\documentclass[aps,pra,amsmath,amssymb,floatfix,twocolumn,amsmath,superscriptaddress,twocolumn,nofootinbib,tighten,letterpaper]{revtex4-2}
\usepackage[colorlinks,linkcolor=blue,citecolor=blue,urlcolor=blue]{hyperref}
\usepackage{multirow}
\usepackage{subfigure}
\usepackage{color}
\usepackage{mathrsfs}
\usepackage{hyperref}
\usepackage[normalem]{ulem}
\usepackage{bm}

\usepackage{amssymb}
\usepackage{amsmath}

\usepackage{tabularray}

\usepackage{array} 
\newcolumntype{P}[1]{>{\centering\arraybackslash}p{#1}}

\definecolor{RowColor}{rgb}{0.88,1,0.9}

\usepackage{amsfonts, relsize, color, mathtools, physics}
\usepackage{graphics}
\usepackage{graphicx}
\usepackage{subfigure}
\usepackage{color}
\usepackage{comment}

\begin{document}
\title{Strained hyperbolic Dirac fermions: Zero modes, flat bands, and competing orders}

\author{Christopher A. Leong}
\affiliation{Department of Physics, Lehigh University, Bethlehem, Pennsylvania, 18015, USA}

\author{Bitan Roy}
\affiliation{Department of Physics, Lehigh University, Bethlehem, Pennsylvania, 18015, USA}

\date{\today}

\begin{abstract}
Starting from the nearest-neighbor tight-binding model on half-filled bipartite hyperbolic $\{ 10, 3\}$ and $\{ 14,3\}$ (Schl\"afli symbols) lattices that, for a uniform hopping amplitude, gives rise to emergent Dirac quasiparticles on a curved space with a constant negative curvature, displaying the characteristic linearly vanishing density of states, we here propose spatially modulated hopping patterns therein that preserve the underlying $5$-fold and $7$-fold rotational symmetries, respectively, and effectively couples such quasi-relativistic fermions to time-reversal symmetric axial magnetic fields. Presence of such strain-induced axial fields, produces a flat band near zero-energy, triggering the nucleation of a charge density-wave, featuring a staggered pattern of fermionic density between complementary sublattices, and the Haldane phase, fostering intra-sublattice circulating currents with a net zero magnetic flux in the system, for subcritical nearest- and next-nearest-neighbor Coulomb repulsions, respectively. Sufficiently weak on-site Hubbard repulsion, on the other hand, destabilizes such axial flat bands toward the formation of a peculiar magnetic phase that simultaneously supports antiferromagnetic and ferromagnetic orders in the whole system. While the magnetization in the bulk and boundary of the system cancel each other, the Ne\'el order continues to be of the same sign everywhere, thereby yielding an edge-compensated global antiferromagnet. Throughout, we draw parallels between these findings and the well-studied qualitatively similar results on a $3$-fold rotationally symmetric strained honeycomb lattice, thereby unifying the phenomenon of axial magnetic catalysis for Dirac fermions, encompassing the ones living on the flat Euclidean and curved hyperbolic planes. Finally, we show that with a specific class of non-Hermiticity, resulting from a balanced gain and loss and manifesting via an imbalance in the hopping amplitudes between the complementary sublattices in the opposite directions, magnitudes of all these orders can be boosted substantially when all the eigenvalues in the noninteracting systems are real, staging a non-Hermitian amplification of axial magnetic catalysis.                     
\end{abstract}

\maketitle

\section{Introduction}~\label{sec:intro} 

Interacting Dirac fermions constitute an ideal platform to showcase emergent relativistic quantum critical phenomena near semimetal-to-insulator and semimetal-to-superconductor quantum phase transitions, respectively realized for strong repulsive and attractive interactions in both flat Euclidean and curved hyperbolic planes~\cite{DiracCriticalityFT:1, DiracCriticalityFT:2, DiracCriticalityFT:3, DiracCriticalityFT:4, DiracCriticalityFT:5, DiracCriticalityFT:6, DiracCriticalityFT:7, DiracCriticalityFT:8, DiracCriticalityFT:9, DiracCriticalityFT:10, DiracCriticalityFT:11}. Since the honeycomb~\cite{DiracLattice:1} and some specific bipartite hyperbolic~\cite{DiracLattice:2} lattices accommodate massless Dirac fermions near half-filling, field theoretic predictions can also be tested independently from lattice-based numerical simulations~\cite{DiracCriticalityNum:1, DiracCriticalityNum:2, DiracCriticalityNum:3, DiracCriticalityNum:4, DiracCriticalityNum:5, DiracCriticalityNum:6, DiracCriticalityNum:7}. However, tuning the requisite interaction-to-bandwidth ratio experimentally in real quantum materials is often challenging to trigger such relativistic Mott or superconducting transitions~\cite{{DesignerDiracCrit}}.

One realistic approach to circumvent this practical limitation (besides applying external strong magnetic fields that, on the other hand, breaks the time-reversal symmetry $\mathcal T$) is to place Dirac materials in axial or pseudo magnetic fields that result from rotational symmetry preserving spatially modulated hopping amplitudes. When a finite flux of such a $\mathcal T$-preserving axial field pierces Dirac materials, a near-zero energy flat band develops in the system, wherein the number of states is proportional to the number of axial flux quanta enclosed by the system. Such a outcome is a direct generalization of the Aharonov-Casher index theorem for axial fields~\cite{index:1, index:2, index:3}, which recently has been extended to hyperbolic Dirac fermions subject to only real magnetic fields, however~\cite{index:4}.

In this work, we unfold a microscopic realization of axial magnetic fields within the framework of the canonical nearest-neighbor (NN) hopping Hamiltonian on specific hyperbolic lattices with spatially modulated hopping amplitudes that in turn couple with massless Dirac fermions and produce flat bands near zero-energy. We show that such flat bands can trigger nucleation of various competing orders even for sufficiently weak local four-fermion interactions, a phenomenon we name \emph{axial magnetic catalysis}. Furthermore, we show that a certain type of system-to-environment coupling (yielding non-Hermiticity in the system) can further boost such ordering tendencies, a phenomenon we call \emph{non-Hermitian amplification of axial magnetic catalysis}. In these contexts, next we present a synopsis of our central findings. 

\begin{figure*}[t!]
\includegraphics[width=0.80\linewidth]{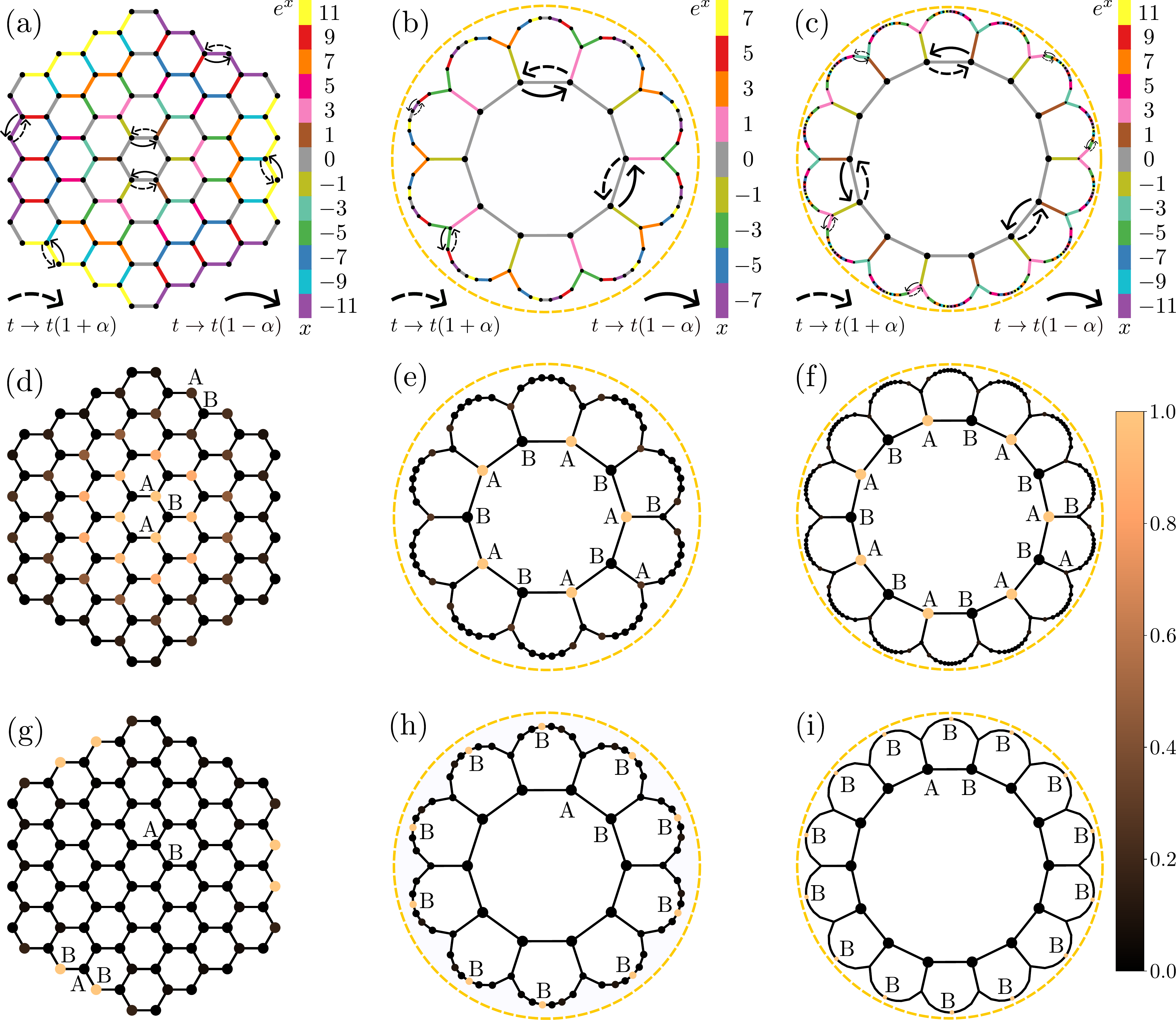}
\caption{Realizations of a strained (a) Euclidean honeycomb lattice, and strained (b) $\{10,3\}$ and (c) $\{14,3\}$ hyperbolic lattices producing time-reversal symmetric axial magnetic fields. The modified hopping amplitudes along the nearest-neighbor (NN) bonds are color coded, showing the preserved three-fold, five-fold, and seven-fold rotational symmetries, respectively. In a non-Hermitian setup, parameterized by a real $\alpha$, the hopping amplitude ($t$) along each NN bond becomes $t(1-\alpha)$ (solid arrow) and $t(1+\alpha)$ (dashed arrow) in the opposite directions marked by the arrow heads between two sublattices ($A$ and $B$), yielding all-real eigenvalues when $|\alpha|<1$. Local density of states (LDOS) associated with the symmetric combination of two particle-hole symmetric closest to zero-energy modes on (d) honeycomb, and (e) $\{ 10, 3 \}$ and (f) $\{ 14, 3\}$ hyperbolic lattices, showing its strong bulk localization on the sites of one sublattice, namely $A$. LDOS associated with the anti-symmetric combination of two particle-hole symmetric closest to zero-energy modes on (g) honeycomb, and (h) $\{ 10, 3 \}$ and (i) $\{ 14, 3\}$ hyperbolic lattices, showing its strong boundary localization on the sites of the other sublattice, namely $B$. Two hyperbolic lattices are always shown on the Poincar\'e disk (golden dashed ring). For details see Secs.~\ref{sec:strainconstruct}, \ref{sec:zeromodes}, and \ref{sec:NHconstruct}.  
}~\label{fig:Geometry}
\end{figure*}

\subsection{Summary of main results}

\begin{figure*}[t!]
\includegraphics[width=0.95\linewidth]{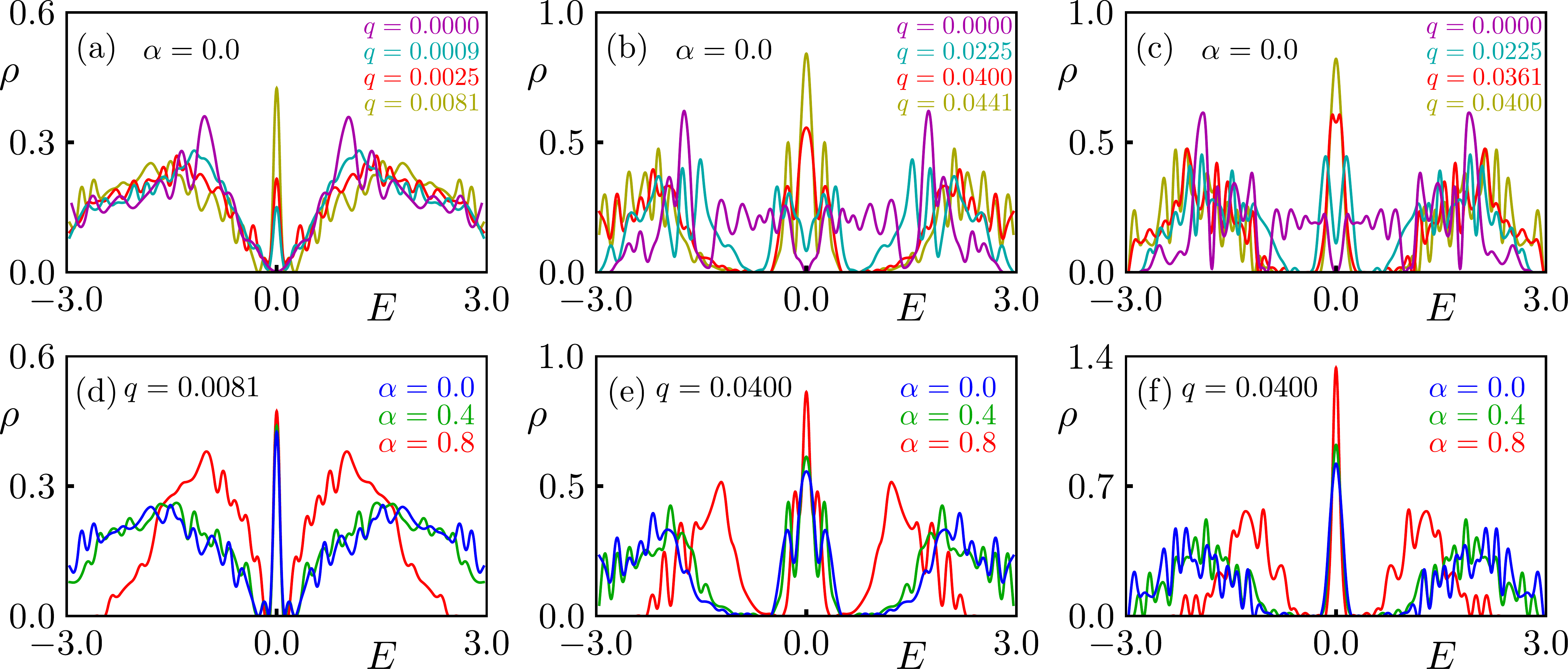}
\caption{Density of states (DOS) $\rho$ as a function of energy $E$ in the absence ($q=0$) and presence (finite $q$) of axial magnetic fields on a (a) honeycomb lattice, and on (b) $\{10,3\}$ and (c) $\{ 14, 3\}$ hyperbolic lattices. Notice that for $q=0$, in all these systems the DOS near zero-energy vanishes linearly with $E$ confirming the presence of massless Dirac fermions therein. With finite $q$, a peak in the DOS develops near $E=0$, yielding a flat band. These results are shown in Hermitian systems ($\alpha=0$). Now introducing the non-Hermitian degree of freedom, the peak in the DOS near $E=0$ gets taller with increasing non-Hermiticity $\alpha$ (see Fig.~\ref{fig:Geometry}) as shown for the (d) honeycomb lattice, and (e) $\{ 10, 3\}$ and (f) $\{ 14,3 \}$ hyperbolic lattices with a fixed nonzero $q$ (axial field) in each system. For details see Sec.~\ref{sec:zeromodes} and Sec.~\ref{sec:NHconstruct}. DOS in pristine honeycomb lattice ($q=0$) is computed with periodic boundary conditions in all directions to suppress the signatures of zero energy topological edge modes, localized near the zigzag boundaries, yielding a finite DOS near $E=0$. 
}~\label{fig:DOS}
\end{figure*}

Within the framework of the NN tight-binding model in two representative $\{ 10, 3\}$ and $\{ 14,3\}$ (Schl\"afli symbol, defined in Sec.~\ref{sec:systemsize}) bipartite hyperbolic lattices, yielding massless Dirac excitations near the half-filling or zero-energy on a curved space with a constant negative curvature, showing the characteristic linearly vanishing density of states (DOS) in the pristine condition, we propose spatially modulated 5-fold and 7-fold rotational symmetric real hopping patterns therein, respectively, that mimic lattice-regularized ${\mathcal T}$-preserving axial magnetic fields (Fig.~\ref{fig:Geometry}). The presence of such a strain-induced axial field produces position-dependent background negative curvature and its coupling with Dirac excitations yields a reconstructed DOS, featuring a prominent flat band near zero energy with the number of states therein increasing with the axial field strength (Fig.~\ref{fig:DOS}).

In contrast to the situation in pristine half-filled hyperbolic Dirac fluids, when strained, these systems become susceptible to the nucleation of various ordered phases for \emph{subcritical} strengths (in comparison to the ones for zero axial field~\cite{biorthogonalQM:2}, see Table~\ref{tab:criticalvalues}) of the finite range components of electronic repulsion due to the presence of such axial flat bands. The exact nature of the ground state with spontaneously broken symmetries depends on their relative strength. Subscribing to the appropriate mean-field theories, we show that the NN and next-nearest-neighbor (NNN) Coulomb repulsions respectively foster a charge-density-wave (CDW) order, see Fig.~\ref{fig:CDW}, and the Haldane order (HO), see Fig.~\ref{fig:Haldane}. While the former displays a staggered pattern of fermionic density between two emergent sublattices of the $\{10,3\}$ and $\{ 14,3 \}$ hyperbolic lattices, the HO stems from intra-sublattice currents~\cite{haldane:1}, so that no net magnetic flux pierces through the system.

Inclusion of the on-site Hubbard repulsion stabilizes a peculiar magnetic state which simultaneously displays both the antiferromagnet (AFM) and ferromagnet (FM) orders, see Figs.~\ref{fig:AFM}-\ref{fig:GlobalAFM143}. However, the net magnetization in the ground state is zero, as the FM order is of opposite sign in the bulk and near the boundary of the system. On the other hand, the AFM Ne\'el order is of the same sign in the entire system, yielding an unconventional magnetic phase which we name edge-compensated global antiferromagnet, solely resulting from the unusual spatial localization of the near zero-energy modes which reside on two complementary sublattices near the bulk and boundary of the system, a feature that uniquely develops in these system due to the presence of axial fields (Fig.~\ref{fig:Geometry}). Extensive finite-size scaling analysis strongly suggests that nucleation of these orders for the subcritical strength of appropriate local four-fermion interaction can be observed in the entire system in the thermodynamic limit (Fig.~\ref{fig:finsizeHer}).

Throughout we draw parallels between these findings and the well-studied honeycomb or $\{ 6,3 \}$ lattice subject to 3-fold symmetric tri-axial strains (also yielding axial fields)~\cite{StrainExpGra:1, StrainExpGra:2, StrainExpGra:3}, where all the above mentioned conclusions have already been established in the literature~\cite{index:3, AxialCataGra:1, AxialCataGra:2, AxialCataGra:3, AxialCataGra:4, AxialCataGra:5, AxialCataGra:6, AxialCataGra:7}. All the results on such strained honeycomb lattice are, however, derived here independently, which nonetheless unfold a number of fascinating features that have remained unnoticed so far, discussed later in the manuscript in detail, see Figs.\ref{fig:Geometry}-\ref{fig:finsizeHer}. Most importantly, these findings altogether bring the emergent phenomena in strained Euclidean and hyperbolic Dirac fluids (in the presence and absence of four-fermion interactions), modeled by the coupling between quasi-relativistic fermions with ${\mathcal T}$-preserving axial fields, under the same umbrella, thereby unifying the findings, summarized above.

Finally, we extend the jurisdiction of the axial magnetic catalysis to a class of non-Hermitian (NH) strained Euclidean and hyperbolic Dirac fluids, in which the system-to-environment coupling manifests via a balanced gain and loss, thereby yielding an all-real eigenvalue spectrum over an extended NH parameter regime. In our specific construction, such a non-Hermiticity results from an imbalance in the NN hopping amplitudes between the complementary sublattices in the opposite directions in the free fermion systems (Fig.~\ref{fig:Geometry}). In such open quantum systems, all the eigenvalues (when real) get squeezed toward zero energy, thus reducing the bandwidth (Fig.~\ref{fig:DOS}). As a result, the magnitude of the CDW, HO, and global AFM gets amplified with increasing strength of the non-Hermiticity, as all of them belong to a special family of orders named commuting class masses (define later in Sec.~\ref{sec:NHamplify}) that fully anticommute with the lattice-regularized NH strained Dirac operator, and hence stand as mass orders therein (Figs.~\ref{fig:CDW}-\ref{fig:AFM}). We coin this phenomenon as non-Hermitian amplification of axial magnetic catalysis, qualitatively similar to the one recently shown for strong external magnetic fields in these systems~\cite{NHampl:RealB}. Adequate finite size scaling analyses strongly suggest that these conclusions remain operative in the thermodynamic limit (see Fig.~\ref{fig:finsizeNH}).

\begin{table}[t!]
\centering
{\renewcommand{\arraystretch}{1.2}
\begin{tabular}{|P{1.6cm}|P{1cm}||P{1.1cm}|P{1.1cm}|P{1.1cm}|}
        \hline
        \multicolumn{1}{|c|}{} & \multicolumn{1}{c||}{} & \multicolumn{3}{c|}{Lattice}\\
        \cline{3-5}
        \vspace{-0.55cm}Interaction & \vspace{-0.55cm}$\alpha$ & $\{6,3\}$ & $\{10,3\}$ & $\{14,3\}$\\
        \hline \hline
        & $0.0$ & 0.695 & 0.670 & 0.652\\
        \cline{2-5}
        & $0.4$ & 0.637 & 0.617 & 0.595\\
        \cline{2-5}
        \vspace{-0.73cm}$V_1$ & $0.8$ & 0.420 & 0.403 & 0.387\\
        \hline
        & $0.0$ & 1.26 & 2.19 & 2.22\\
        \cline{2-5}
        & $0.4$ & 1.19 & 1.96 & 2.01\\
        \cline{2-5}
        \vspace{-0.73cm}$V_2$ & $0.8$ & 0.83 & 1.30 & 1.32\\
        \hline
        & $0.0$ & 2.08 & 1.46 & 1.34\\
        \cline{2-5}
        & $0.4$ & 1.91 & 1.34 & 1.24\\
        \cline{2-5}
        \vspace{-0.73cm}$U$ & $0.8$ & 1.24 & 0.87 & 0.79\\
        \hline
\end{tabular}
}
\caption{Critical values of the nearest-neighbor Coulomb ($V_1$), next-nearest-neighbor Coulomb ($V_2$), and on-site Hubbard ($U$) repulsions for the Euclidean ($\{6,3\}$) and hyperbolic ($\{10,3\}$ and $\{14,3\}$) Dirac systems with zero ($\alpha=0.0$) and finite ($\alpha=0.4$ and $0.8$) amounts of non-Hermiticity, in the absence of any axial magnetic field for the charge density-wave, Haldane order, and antiferromagnet, respectively, in the mean-field limit. See Ref.~\cite{biorthogonalQM:2} and Appendix~\ref{app:nostrainhaldane}.
}~\label{tab:criticalvalues}
\end{table}

\subsection{Organization}

The remainder of the paper is organized as follows. In Sec.~\ref{sec:systemsize}, we identify the Euclidean and hyperbolic lattices yielding Dirac fermions at half-filling in pristine condition, their dimensions (in terms of the number of lattice sites), and the boundary conditions, imposed therein. A modified hopping model mimicking axial magnetic fields on Euclidean and hyperbolic Dirac lattices is discussed in Sec.~\ref{sec:strainconstruct}. In Sec.~\ref{sec:zeromodes}, we discuss the structure of the zero-energy modes, constituting the axial flat bands. Analytical arguments in favor of axial magnetic catalysis is presented in Sec.~\ref{sec:axialmagcat}. Effects of NN ($V_1$) and NNN ($V_2$) Coulomb interactions on spinless fermions and axial catalysis of the resulting CDW and HO are discussed in Secs.~\ref{sec:coulomb} and~\ref{sec:haldane}, respectively. The effects of the on-site Hubbard repulsion ($U$) on the axial flat bands and the resulting edge-compensated global AFM is discussed in Sec.~\ref{sec:hubbard}. Construction of the NH tight-binding model in the presence of axial field or strain is introduced in Sec.~\ref{sec:NHconstruct}. Section~\ref{sec:NHamplify} is devoted to promote the notion of NH amplification of the axial catalysis mechanism. The requisite tools of the biorthogonal quantum mechanics for the self-consistent numerical simulations of NH systems are briefly reviewed in Sec.~\ref{sec:biorthogonal}. In Sec.~\ref{sec:NHinteract}, we discuss the numerical results endorsing the NH amplification of axial catalysis. Finite size effects of all our results are discussed in Sec.~\ref{sec:finitesize}. Summary of results and discussions on related topics, including future directions and possible paths for experimental verifications of our theoretical results, are presented in Sec.~\ref{sec:summarydiscussion}. In the only appendix (Appendix~\ref{app:nostrainhaldane}), we compute the critical strengths of the NNN Coulomb repulsion for the HO in the absence of any axial field, but in both Hermitian and NH, Euclidean and hyperbolic Dirac liquids.

\section{System specification}~\label{sec:systemsize}

A two-dimensional crystal is described by a periodic arrangement of regular polygons that is translationally invariant and features various discrete crystallographic symmetries (such as rotations), depending on its connectivity. Accordingly, these systems are characterized by the $p$-sided polygon that tiles the two-dimensional plane and the coordination number $q$ of each vertex, which is expressed compactly using the Schl\"afli symbol $\{p, q\}$. On a flat Euclidean surface, periodic tilings are only allowed when $(p-2)(q-2)=4$, the solutions of which are exhausted by $2p=q=6$ (triangle lattice), $p=q=4$ (square lattice), and $p=2q=6$ (honeycomb lattice). By contrast, tilings on a hyperbolic space must conform to $(p-2)(q-2)>4$, of which there exists an infinite number of solutions and due to a background constant negative curvature, the resulting hyperbolic lattices can display a plethora of peculiar physical outcomes that have attracted ample attention in recent time~\cite{HL:1, HL:2, HL:3, HL:4, HL:5, HL:6, HL:7, HL:8, HL:9, HL:10}.

On the other hand, when considering the free fermion tight-binding model with the canonical NN hopping, only the honeycomb lattice and any hyperbolic lattice with $p/2$ belonging to the set of all odd integers and $q=3$ give rise to massless Dirac fermions at half-filling, with vanishing density of states (DOS), therein. Therefore, in this work we only consider the Euclidean honeycomb or $\{6,3\}$, and $\{ 10,3\}$ and $\{14,3\}$ hyperbolic lattices, yielding massless Dirac fermions in flat Euclidean and  curved (with constant negative curvature) spaces, respectively.

Throughout, on all these lattices, the center plaquette corresponds to the zeroth generation, and each successive layer of plaquettes constitutes its progressively next generation. Unless mentioned otherwise, all the calculations are performed on honeycomb lattices with nineteen generations, containing 2400 sites, and $\{10,3\}$ and $\{14,3\}$ hyperbolic lattices with three and two generations, respectively, containing 2880 and 1694 sites.

All the numerical results are obtained from exact numerical diagonalization of real space tight-binding Hamiltonian in noninteracting systems and effective single-particle Hamiltonian in interacting systems with open boundary conditions (unless mentioned otherwise). The latter ones are obtained after decomposing the four-fermion NN and on-site Hubbard repulsions in the Hartree channel, and NNN repulsion in the Fock channel.

\begin{figure*}[t!]
\includegraphics[width=0.95\linewidth]{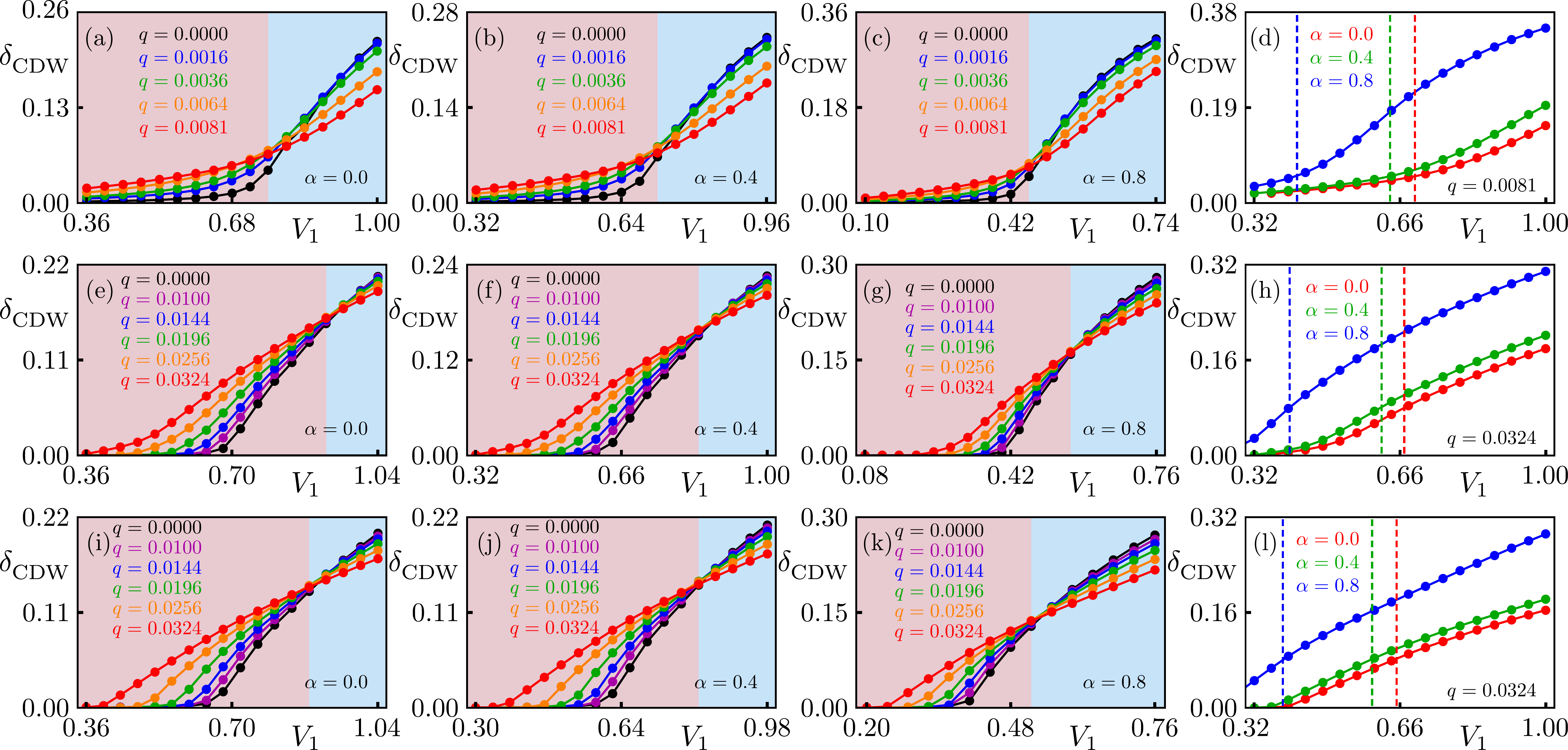}
\caption{Scaling of the self-consistent solutions of the charge density wave (CDW) order, averaged over the entire system ($\delta_{\rm CDW}$), as a function of the nearest-neighbor Coulomb repulsion ($V_1$) among spinless fermions on a honeycomb lattice in a (a) Hermitian setup ($\alpha=0$) and in non-Hermitian setups with (b) $\alpha=0.4$ and (c) $\alpha=0.8$ (see Fig.~\ref{fig:Geometry}) in the absence of any axial magnetic field ($q=0$) and for a few choices of finite $q$ (producing axial magnetic fields). While for $q=0$, $\delta_{\rm CDW}$ strictly develops beyond an $\alpha$-dependent critical strength of $V_1$, denoted by $V_{1,c}$, see Table~\ref{tab:criticalvalues}, finite axial magnetic fields catalyze their nucleation for subcritical $V_1$ due to the presence of a flat band near zero energy (see Fig.~\ref{fig:DOS}). As finite $q$ also increases the bandwidth of the free fermion system, we observe a crossover between a strain-dominated regime (red shade) for weaker $V_1$ where a larger $q$ yields a bigger $\delta_{\rm CDW}$ and an interaction-dominated regime (light blue shade) for stronger $V_1$ where a larger $q$ yields a smaller $\delta_{\rm CDW}$. Although this is a crossover phenomenon, all the data for $\delta_{\rm CDW}$ for various finite $q$ seem to cross at a specific $V_1 > V_{1, {\rm c}}$. Amplification of the axial catalysis of the CDW order by the non-Hermiticity in the system for any $V_1$ is shown in (d) by comparing $\delta_{\rm CDW}$ for a fixed $q$ (axial field) but for $\alpha=0.0$, $0.4$, and $0.8$ over a wide range of $V_1$. Notice that for a fixed $q$ and $V_1$, $\delta_{\rm CDW}$ increases monotonically with $\alpha$. Vertical lines in (d), color coded according to the $\alpha$ values, mark the $\alpha$-dependent critical $V_1$ for the CDW ordering when $q=0$. Panels (e)-(h) [(i)-(l)] are analogous to panels (a)-(d), respectively, but for the $\{10,3 \}$ [$\{14,3 \}$] hyperbolic Dirac lattice. See Sec.~\ref{sec:coulomb} and Sec.~\ref{sec:NHinteract} for detailed discussions.   
}~\label{fig:CDW}
\end{figure*}

\section{Axial magnetic fields}~\label{sec:strainconstruct}

We begin this section by considering the NN tight-binding model on the honeycomb or $\{6,3\}$, $\{10,3\}$, and $\{14,3\}$ lattices. In pristine condition, all of these systems host massless Dirac fermions at half-filling, displaying the characteristic linearly vanishing DOS near zero energy (shown in Fig.~\ref{fig:DOS})~\cite{DiracLattice:1, DiracLattice:2}. The model also gives rise to a microscopic (on $\{6,3\}$ lattice) or an emergent (on $\{ 10,3\}$ and $\{14,3\}$ lattices) bipartite structure (since $p$ is always an even integer), thereby presenting a sublattice exchange symmetry between two sublattices. Upon introducing a defect, described by buckling the 2D sheet of the lattice into a bulge deformation, the bonds between sites are deformed, resulting in a modulation of the hopping amplitudes shown in Fig.~\ref{fig:Geometry} and is described by the Hamiltonian
\allowdisplaybreaks[4]
\begin{equation}~\label{eq:hstrain}
     \begin{split}H_{\rm strain} &= \sum\limits_{\langle a,b \rangle} e^{\chi(R_a)}t_{a,b}e^{-\chi(R_b)}c_{b}^{\dagger}c_a^{} + {\rm H.c.}\\
    &=\psi^\dagger\begin{pmatrix}
        \mathbf{0} & \mathbf{t_{\rm strain}}\\
        \mathbf{t_{\rm strain}^{\top}} & \mathbf{0}
    \end{pmatrix}\psi \equiv \psi^\dagger \hat{h}_{\rm strain} \psi.
    \end{split}
\end{equation}
Here the summation is restricted over NN sites denoted by $\langle\dots\rangle$, $a$ ($b$) is an index running over all the sites in the $A$ ($B$) sublattice, $t_{a,b}$ is the hopping amplitude between sites $a$ and $b$, $c_b^{\dagger}$ ($c_a^{}$) is the creation (annihilation) operator acting on site $b$ ($a$), $\chi$ is a function whose form depends on the profile of the resulting strain-induced axial magnetic field, $R_a$ ($R_b$) is the distance of site $a$ ($b$) from the center of the bulge, ${\rm H.c.}$ denotes the Hermitian conjugate, and the spinor basis of the block matrix is $\psi^\dagger = (c_A, c_B)^\dagger$ with $c_A$ ($c_B$) as column spinor containing the annihilation operators in the sites of $A$ ($B$) sublattice. The off-diagonal block matrix form of the operator owes to the bipartite structure, with $\mathbf{t_{\rm strain}}$ representing the inter-sublattice hopping in the presence of the defect or strain. This construction is shown explicitly on the $\{6,3\}$, $\{10,3\}$, and $\{14,3\}$ lattices in Figs.~\ref{fig:Geometry}(a),~\ref{fig:Geometry}(b), and~\ref{fig:Geometry}(c), respectively. On the Euclidean plane, a uniform axial magnetic field is realized by $\chi(R)=qR^2$, resulting in a strength of the field equal to $2q$, where $q$ is a \emph{real} tunable parameter. In this work we only consider this form of the axial magnetic field for both the Euclidean and hyperbolic lattices. Notice that this construction preserves the microscopic $p/2$-fold rotational symmetry of the underlying lattices. Additionally, as a consequence of the hopping deformation, the Euclidean honeycomb lattice develops a spatially dependent negative curvature, while on the hyperbolic lattice, already embedded on a negatively curved space, the curvature becomes spatially dependent.

The physical consequence of the axial magnetic field is the formation of Landau levels, with the most important being the one sitting at zero energy. The emergence of these zero modes, which would otherwise be absent in an unstrained Dirac system, allow for the catalysis of dynamic mass generation even at sufficiently weak coupling, thereby manifesting the axial magnetic catalysis mechanism (discussed in detail in Sec.~\ref{sec:axialmagcat}).

\section{Zero Modes}~\label{sec:zeromodes}

We have outlined the modified tight-binding Hamiltonian in Euclidean and hyperbolic Dirac systems due to strain, see Fig.~\ref{fig:Geometry}(a)-\ref{fig:Geometry}(c), which gives rise to near zero energy modes. In this section, we focus on the behavior of these modes in systems devoid of interactions. This is accomplished by studying the DOS and local DOS, obtained from exact diagonalization of the Hamiltonian given in Eq.~\eqref{eq:hstrain} for the strained honeycomb or $\{6,3\}$, $\{10,3\}$, and $\{14,3\}$ lattices, and subsequent binning of the eigenvalues within different energy windows.

We begin this discussion by considering the zero-energy eigenvectors, which are given by~\cite{AxialCataGra:4, AxialCataGra:5}
\begin{equation}
    \Psi_{\rm strain}(0) = \exp[-\chi\; \text{diag.}(\mathbf{I}, \mathbf{-I})]\Psi_{\rm strainless}(0),
    \label{eq:strainwavefunction}
\end{equation}
where $\Psi_{\rm strain}(0)$ [$\Psi_{\rm strainless}(0)$] is the wavefunction of the system in the presence [absence] of strain for zero energy and $\mathbf{I}$ is an $N/2$ dimensional identity matrix in the same spinor basis $\psi$ defined in the last section. In this representation, it can clearly be seen that the $A$ and $B$ sublattices host normalizable and unnormalizable wavefunctions, respectively. Consequently, in a system with open boundary conditions, the former is more localized in the bulk, while the latter is localized on the open boundary edges. We seek to study this localization effect on finite sized lattice systems, however then there are no exact zero energy modes. Therefore, to numerically study this phenomenon, we identify the two closest to zero energy states, given by $\ket{\delta^+}$ and $\ket{\delta^-}$, with energies $\delta^+$ and $\delta^- (=-\delta^+)$, and form their symmetric (S) and antisymmetric (AS) combinations, respectively given by 
\begin{equation}
    \ket{\rm S} = \frac{\ket{\delta^+} + \ket{\delta^-}}{\sqrt{2}} 
    \: \text{and} \: 
    \ket{\rm AS} = \frac{\ket{\delta^+} - \ket{\delta^-}}{\sqrt{2}}.
\end{equation}

With these states in hand, we compute their LDOS, which is shown in Fig.~\ref{fig:Geometry}(d),~\ref{fig:Geometry}(e), and~\ref{fig:Geometry}(f) [Fig.~\ref{fig:Geometry}(g),~\ref{fig:Geometry}(h), and~\ref{fig:Geometry}(i)] for the symmetric [antisymmetric] state on the $\{6,3\}$, $\{10,3\}$, and $\{14,3\}$ lattices, respectively. These results show that all systems display localization of the near zero energy modes $\ket{\rm S}$ on the $A$ sublattice in the bulk and $\ket{\rm AS}$ on the $B$ sublattice on the edge, as predicted in the continuum limit. It should, however, be noted that we have symmetrized the states in order to guarantee the expected crystallographic symmetries. This step is needed because a numerical eigensolver does not necessarily yield eigenstates of the rotational symmetry operator. Therefore, if these non-symmetrized eigenvectors were to be used, physical observables, such as the LDOS, would seemingly favor one region of the lattice over the other even when two such regions are connected by a crystallographic symmetry. Such a result is not physical and thus needs to be corrected via symmetrization.

Having discussed the local DOS of the near zero-energy modes, now we focus on the DOS near zero energy as a function of the axial magnetic field strength ($q$). Studying first the honeycomb or $\{6, 3\}$ lattice, displayed in Fig.~\ref{fig:DOS} (a), we observe, in the presence of a strong axial magnetic field, a sharp spike in the DOS at zero energy, capturing the formation of the zeroth axial Landau level. Moreover, with increasing strength of the field, we note that the number of zero energy modes increases, consistent with the Aharonov-Casher index theorem, generalized to axial magnetic fields and negatively curved space. All of these observations remain operative in hyperbolic systems, displayed in Fig.~\ref{fig:DOS}(b) and~\ref{fig:DOS}(c) for the $\{10,3\}$ and $\{14,3\}$ lattices, respectively, with a key distinction. In the hyperbolic case, there are no sharp Landau levels, which is possibly due to the background negative curvature, as also observed with real magnetic fields~\cite{realBHypNoLL}. Nonetheless, the connection between the axial magnetic field and the formation of zero energy modes has been unequivocally established here. We also note that with increasing strength of the axial field, or equivalently $q$, the DOS near zero energy increases and concurrently the bandwidth of the system increases as well (due to a stronger lattice deformation). Such competing outcomes leave their clear signatures on the scaling behavior of various orderings at weak and strong couplings, respectively dominated by the DOS near zero-energy and the bandwidth, which we discuss in subsequent sections.

\begin{figure*}[t!]
\includegraphics[width=0.95\linewidth]{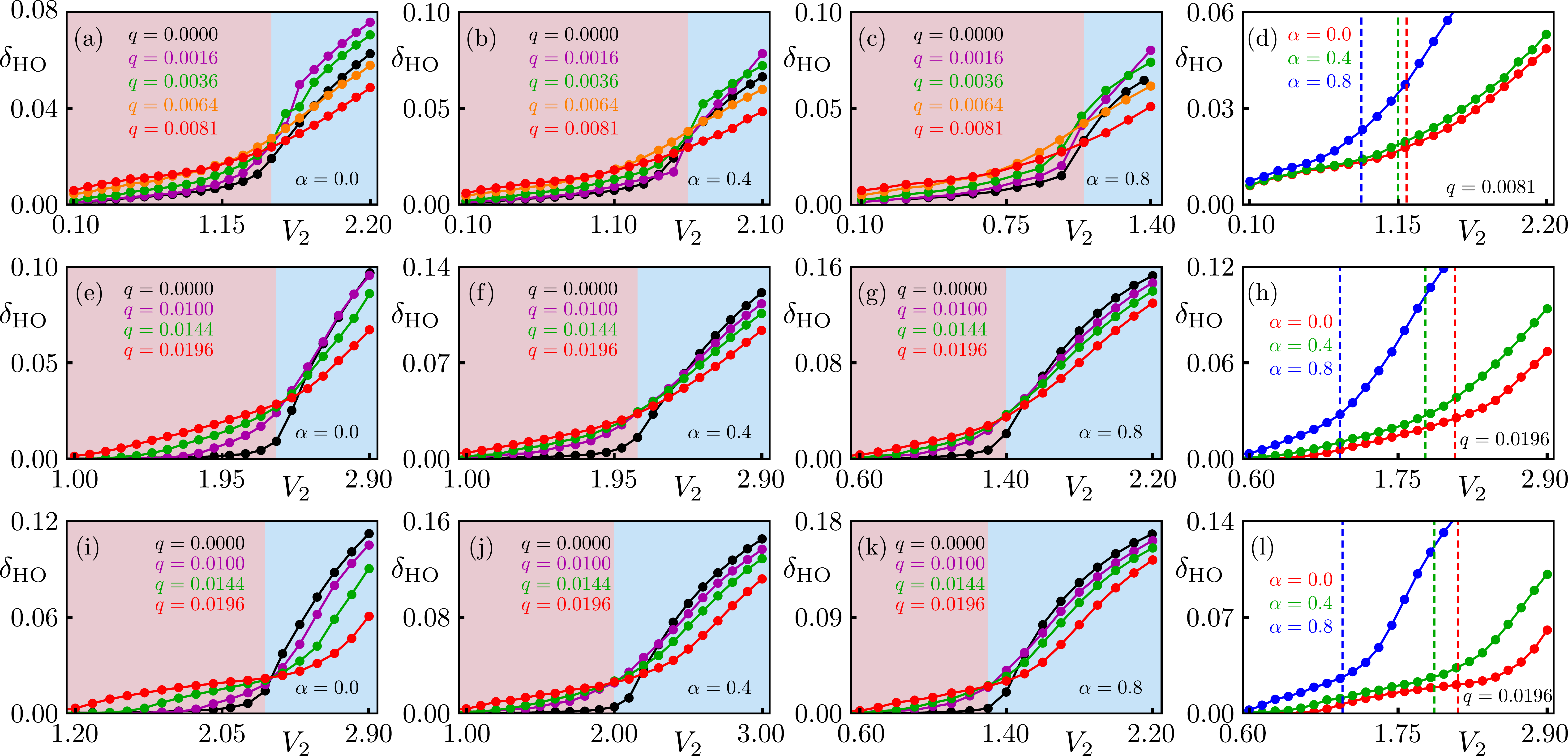}
\caption{Analogous to Fig.~\ref{fig:CDW}, but showing the scaling of the self-consistent solutions of the Haldane order ($\delta_{\rm HO}$), averaged over the entire system, with the next-nearest-neighbor Coulomb repulsion ($V_2$) among spinless fermions. Notice that the anticipated crossing of all data for $\delta_{\rm HO}$ for various $q$ values at the crossover point between the strain-dominated (red shade) and interaction-dominated (light blue shade) regimes at a finite $V_2$ is not perfect for the honeycomb lattice (top row), while such crossing seems to be perfect at a specific $V_2$ on $\{ 10,3\}$ (middle row) and $\{ 14,3\}$ (bottom row) hyperbolic lattices. Such data crossings between the two regions occurs above critical $V_2\; (\equiv V_{2,c})$, which depends on $\alpha$ (see Table~\ref{tab:criticalvalues}). For details see Secs.~\ref{sec:haldane} and~\ref{sec:NHinteract}.   
}~\label{fig:Haldane}
\end{figure*}

\section{Axial Magnetic Catalysis}~\label{sec:axialmagcat}

With the construction of the strain-induced axial magnetic field established and the resulting zero energy modes studied, we now focus on the axial magnetic catalysis mechanism by incorporating various finite range electronic interactions in such systems. To exemplify this phenomenon, we introduce NN Coulomb and on-site Hubbard interactions, respectively giving rise to a staggered patterning of electronic density, called CDW order, and a staggered patterning of electronic spin, called AFM order in half-filled bipartite systems with real hopping amplitudes. Since the latter requires the inclusion of the spin degrees of freedom, we then generalize Eq.~\eqref{eq:hstrain} by including electronic spin yielding the free-fermion Hamiltonian in strained systems $\sigma_0 \otimes \hat{h}_{\rm strain}$, where $\sigma_0 = {\rm diag.}(1,1)$ is the two-dimensional identity matrix and $\otimes$ is the Kronecker product. The set of Pauli matrices $\{ \sigma_\mu \}$ acts on the spin index. These orders represent masses for Dirac systems, as the corresponding matrix operators anticommute with the free-fermion operator $\sigma_0 \otimes \hat{h}_{\rm strain}$, thereby introducing a mass gap into the spectrum near the zero energy via spontaneous symmetry breaking. This can be verified explicitly by considering the operators representing the CDW and AFM orders, which are, respectively, given by
\begin{eqnarray}~\label{eq:massoperators}
\hat{\mathcal O}_{\rm CDW} &=& \sigma_0 \otimes \left( 
\begin{array}{cc}
{\boldsymbol \Delta} & {\boldsymbol 0} \\
{\boldsymbol 0} & -{\boldsymbol \Delta}
\end{array}
\right) \nonumber \\
\text{and} \: \hat{\mathcal O}_{\rm AFM} &=& \sigma_j \otimes \left( 
\begin{array}{cc}
{\boldsymbol \Delta} & {\boldsymbol 0} \\
{\boldsymbol 0} & -{\boldsymbol \Delta}
\end{array}
\right),
\end{eqnarray} 
where $j \in \{1,2,3\}$ and $\boldsymbol{\Delta}$ is an $N/2$-dimensional \emph{real} diagonal matrix, whose entries can be equal or different. In pristine Dirac systems, any finite expectation value of these orderings can only be realized beyond a critical strength of the corresponding interaction, due to their characteristic vanishing DOS.

Now, we establish how the axial magnetic field promotes the nucleation of these orders at sufficiently weak coupling. First, denote the spectrum of the free-fermion model $\sigma_0 \otimes \hat{h}_{\rm strain}$ as $\{E\}$, with $E_i$ denoting the $i$th element in the set. Then, with the introduction of the mass term, described by the effective single-particle Hamiltonian $\sigma_0 \otimes \hat{h}_{\rm strain} + \hat{\mathcal{O}}_j$ for $j= {\rm CDW}$ or AFM, each element of the energy spectrum becomes $E_i \rightarrow E_i^{\Delta} = \sqrt{E_i^2 + \Delta^2}$, where we take the entries of $\hat{\mathcal O}_{\rm CDW}$ or $\hat{\mathcal O}_{\rm AFM}$ to be uniform (namely $\Delta$) for simplicity. Therefore, the free energy of the system in the ground state at zero temperature (all the states at negative energies are filled, while the ones at positive energies are left empty) is given by
\begin{equation}
    F = \frac{\Delta^2}{2g} - D_0 \; |\Delta| - {\sum\limits_{i=1,2, \cdots}}^\prime \; D_i \; E_i^{\Delta},
    \label{eq:freeenergy}
\end{equation}
where $g$ is the coupling constant of the corresponding ordering (such as CDW or AFM), $D_i$ is the degeneracy of the $i$th eigenvalue, and the sum is restricted to negative energies (denoted by the prime). Minimizing the free energy with respect to the magnitude of the mass order ($\Delta$), we obtain the self-consistent gap equation
\begin{equation}
    \frac{1}{g} = \frac{D_0}{|\Delta|} + {\sum\limits_{i=1,2,\cdots}}^\prime \frac{D_i}{\sqrt{E_i^2 + \Delta^2}}.
    \label{eq:freeenergymin}
\end{equation}
Note that even for an infinitesimal $g$ there exists a non-vanishing solution for $\Delta$, given by $\Delta \sim g$ (in this limit the contributions from the second term can be neglected), implying a gap opening up around zero energy. Moreover, with an increased degeneracy of the zeroth axial Landau level (bigger $D_0$), the gap near zero energy becomes increasingly wider. Inclusion of the contribution from the second term further amplifies the self-consistent solution of the ordering amplitude $\Delta$, as any mass ordering, besides splitting the near zero-energy manifold, also pushes all the filled states further down. These conclusions exemplify the axial magnetic catalysis mechanism, through which a Dirac system acquires mass even at sufficiently small coupling strengths.

In addition to the CDW and AFM orders discussed above, we also consider a HO, resulting from the NNN Coulomb repulsion and presenting a topological insulator in graphene at half-filling~\cite{haldane:1, haldane:2}, but not in the half-filled hyperbolic lattices~\cite{haldane:3}. The matrix operator representing the HO is given by
\begin{equation}
    \hat{\mathcal{O}}_{\rm HO} = \sigma_0 \otimes \begin{pmatrix}
        \boldsymbol{t_2} & \boldsymbol{0}\\
        \boldsymbol{0} & \boldsymbol{t_2^{\star}}
        \label{eq:haldaneorder}
    \end{pmatrix},
\end{equation}
where $\boldsymbol{t_2}$ is an $N/2$-dimensional block matrix representing \emph{purely imaginary} NNN hopping and $\star$ denotes the complex conjugation. By construction, the Haldane currents are threaded through the system in such a way that the net magnetic flux through each plaquette of the system is precisely zero. This is encoded by the equal, but complex conjugated intra-sublattice currents imposed on the $A$ and $B$ sublattices. Additionally, notice that $\{\sigma_0 \otimes \hat{h}_{\rm strain}, \hat{\mathcal{O}}_{\rm HO}\} \approx 0$, with the approximation stemming from the finiteness of the systems. Thus the HO also describes a mass, opening up a gap near zero energy. As a result, the above arguments based on the free energy should remain operative for the HO as well. It should be noted that the anticommutation relation between the tight-binding Hamiltonian and the HO is exact in terms of the corresponding Bloch matrix operators.

We note that the CDW order breaks the Isinglike sublattice exchange symmetry. The AFM order, on the other hand, in addition breaks the time-reversal symmetry, generated by $\sigma_2 {\mathcal K}$ where ${\mathcal K}$ is the complex conjugation operator, and the SU(2) spin rotational symmetry, generated by $(\sigma_1, \sigma_2, \sigma_3)$. Finally, the HO breaks the sublattice exchange symmetry and the time-reversal symmetry. Next we set out to establish the axial magnetic catalysis of CDW, HO, and AFM from their numerical self-consistent solutions on strained $\{ 6,3 \}$, $\{ 10,3 \}$, and $\{ 14,3\}$ lattices after appropriate mean-field decompositions of the repulsive NN Coulomb, NNN Coulomb, and on-site Hubbard interactions, respectively.

\section{NN Coulomb repulsions and Charge density-wave}~\label{sec:coulomb}

Having justified the axial magnetic catalysis of the CDW order in the last section, representing a staggered patterning of the fermionic density between the complementary sublattices, here we further support these predictions with numerical computations of strained Dirac systems with NN Coulomb repulsions, yielding the CDW order in the mean-field limit. To begin, the NN Coulomb interaction is included through the Hamiltonian
\begin{equation}
    H_{V_1} = \frac{V_1}{2}\sum\limits_{\langle i,j \rangle}n_in_j - \mu N,
    \label{eq:coulomb}
\end{equation}
where $V_1 (>0)$ is a tunable parameter controlling the strength of the NN Coulomb repulsion, the sum is restricted to NN sites denoted by $\langle...\rangle$, $n_i$ is the fermionic density operator on the $i$th site, and $\mu$ is the chemical potential. Additionally, since there is no coupling to the spin degree of freedom, here we assume spin-polarized or spinless fermions for simplicity. In order to numerically study this interaction in large systems, a Hartree decomposition of $H_{V_1}$ is performed, giving the effective single-particle Hamiltonian~\cite{AxialCataGra:6}
\begin{equation}~\label{eq:coulombmft}
    H_{V_1}^{\rm Har} = V_1 \sum\limits_{\langle i,j \rangle} \left ( \langle n_{A,i} \rangle n_{B,j} + \langle n_{B,i} \rangle n_{A,j} \right ) -\mu N,
\end{equation}
where $\langle n_{A,i} \rangle$ ($\langle n_{B,j} \rangle$) is the average fermionic density of the $i$th ($j$th) site on the $A$ ($B$) sublattice. To maintain the system at half-filling, we set $\mu=V/2$. Now equipped with the full effective single-body Hamiltonian $\hat{h}_{\rm strain} + \hat{h}_{V_1}^{\rm Har}$, we study the emergence of CDW order in the $\{6,3\}$, $\{10,3\}$, and $\{14,3\}$ Dirac systems by self-consistently computing $\langle n_{A,i}\rangle = 1/2 + \delta_{A,i}$ and $\langle n_{B,i}\rangle = 1/2 - \delta_{B,i}$ on each site, where $\delta_{A/B,i}$ is the deviation in average fermionic density from half-filling on the $i$th site of the $A/B$ sublattice, $\delta_{A/B,i}>0$, and $\hat{h}_{V_1}^{\rm Har}$ is the matrix operator associated with $H_{V_1}^{\rm Har}$ from Eq.~\eqref{eq:coulombmft}. The order parameter for the CDW order takes the form 
\begin{equation}
\delta_{\rm CDW} = (\delta_A + \delta_B) / 2,
\end{equation}
where $\delta_A$ ($\delta_B$) is the average value over all $\delta_{A,i}$ ($\delta_{B,i}$) over the entire system. Following this route, we arrive at solutions for $\delta_{\rm CDW}$ as a function of the coupling strength $V_1$ for the honeycomb or $\{6,3\}$, $\{10,3\}$, and $\{14,3\}$ Dirac systems for various choices of $q$, pictured in Fig.~\ref{fig:CDW}(a),~\ref{fig:CDW}(e), and~\ref{fig:CDW}(i), respectively.

Focusing first on subcritical interaction strengths (see Table~\ref{tab:criticalvalues}), we observe that a finite CDW order develops in the system when a sufficiently strong axial magnetic field is present therein, with a greater field strength resulting in an amplified CDW ordering. This observation clearly endorses the axial magnetic catalysis we promoted in Sec.~\ref{sec:axialmagcat} through analytical arguments. Now considering stronger, above-critical strengths of the coupling, we observe the opposite behavior, where stronger axial magnetic fields result in a weaker CDW order. Such contrasting scaling behaviors in the weak (subcritical) and strong (above critical) regimes can be justified in the following way. In the subcritical interaction regime, the ordering is triggered by the near-zero energy manifold whose degeneracy increases with increasing strength of the axial field or $q$, yielding a larger CDW order with increasing field strength. In this regime, the modes far from zero energy play no role. By contrast, in the strong coupling regime, when the interaction strength is greater than the zero field critical one (reported in Table~\ref{tab:criticalvalues}), the role of the near zero-energy modes on the ordering tendency becomes sub-dominant, instead being determined by the bandwidth of the system. Then the increasing bandwidth with larger $q$ or axial field results in a weaker CDW order in the strong coupling regime. We observe a qualitatively similar contrasting behaviors for the HO and AFM in the weak and strong coupling regimes, which we discuss in the next two sections, respectively.

Due to the spatial separation of the near zero-energy modes localized on the $A$ and $B$ sublattices, discussed in Sec.~\ref{sec:zeromodes} and shown in Fig.~\ref{fig:Geometry}, the formation of the CDW order in strained Dirac lattices occurs in the following way. Recall that near zero-energy modes on the $A$ ($B$) sublattices reside in the bulk (near the edges) of the system. Therefore, the CDW in strained Dirac lattices develop by creating a net staggered density between the bulk and boundary of the system, in turn yielding a net CDW order in the entire system~\cite{AxialCataGra:6}.

\begin{figure*}[t!]
\includegraphics[width=0.95\linewidth]{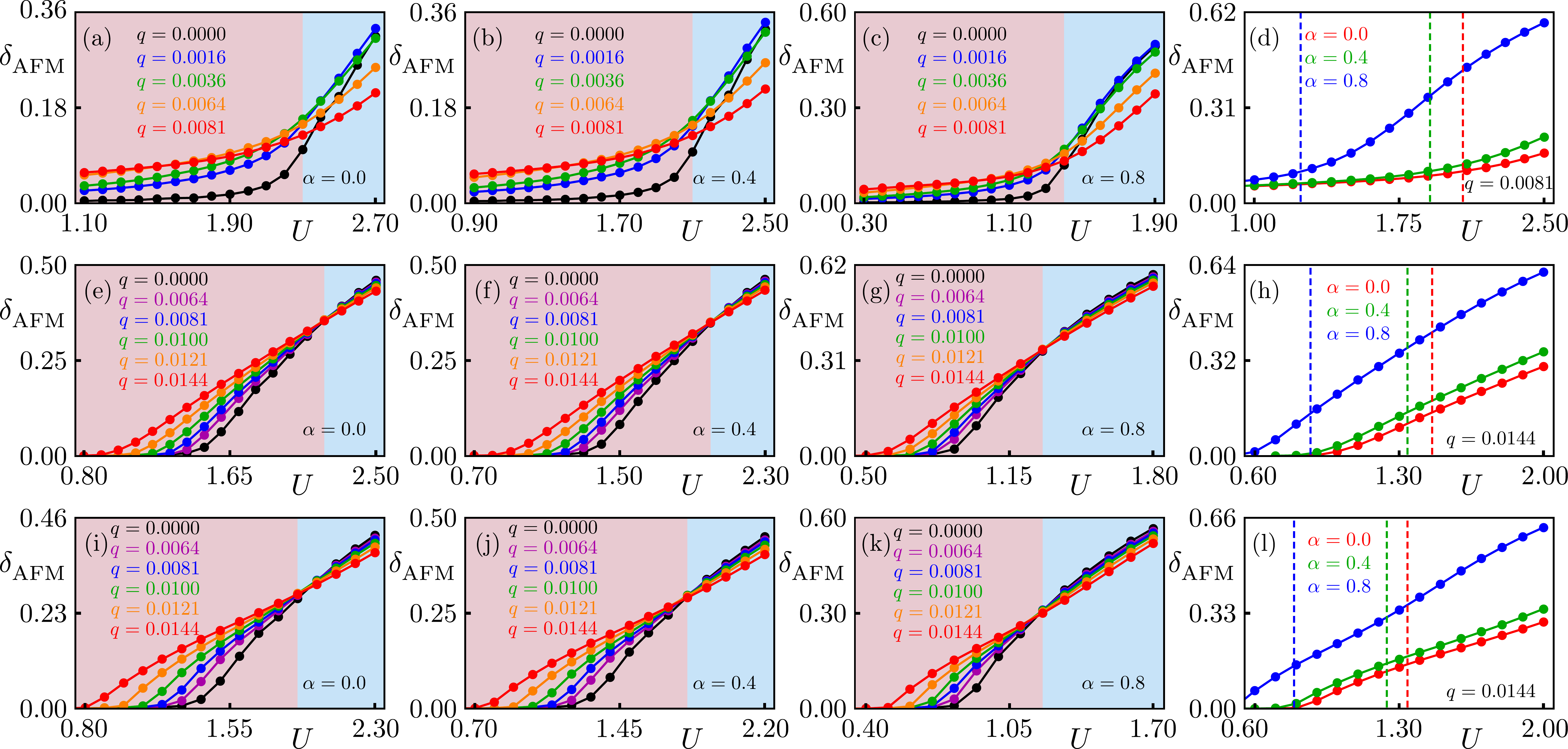}
\caption{Analogous to Figs.~\ref{fig:CDW} and~\ref{fig:Haldane}, but showing the scaling of the self-consistent solutions of the antiferromagnetic order ($\delta_{\rm AFM}$), averaged over the entire system, with the on-site Hubbard repulsion ($U$) among spinful fermions. The anticipated crossing of all data for $\delta_{\rm AFM}$ for various $q$ values at the crossover point between the strain-dominated (red shade) and interaction-dominated (light blue shade) regimes at a finite $U$ is not perfect for the honeycomb lattice (top row), while such crossing seems to be perfect at a specific $U$ on $\{ 10,3\}$ (middle row) and $\{ 14,3\}$ (bottom row) hyperbolic lattices. This data crossing between the two regimes occurs above the critical $U \; (\equiv U_c)$, which depends on $\alpha$ (see Table~\ref{tab:criticalvalues}). For details see Secs.~\ref{sec:hubbard} and~\ref{sec:NHinteract}.   
}~\label{fig:AFM}
\end{figure*}

\section{NNN Coulomb repulsion and Haldane order}~\label{sec:haldane}

The previous section discussed the emergence of the CDW order from subcritical NN Coulomb repulsions. Here, we expand the discussion by incorporating the NNN component of the finite range Coulomb repulsion, which are captured by the Hamiltonian
\begin{equation}
    H_{V_2} = V_2 \sum\limits_{\langle\langle i,j \rangle\rangle} n_in_j -\mu N,
    \label{eq:haldane}
\end{equation}
where $V_2(>0)$ is a tunable parameter controlling the strength of the NNN repulsion, the summation is restricted to NNN sites denoted by $\langle\langle \dots \rangle\rangle$, and $n_i$ is the fermionic density operator acting on site $i$. Additionally, since the NNN repulsion and the resulting HO are spin-independent, we continue to work with spinless or spin-polarized fermions. Decomposing the quartic term now in the Fock channel, we obtain the corresponding effective single-particle Hamiltonian~\cite{AxialCataGra:4} 
\begin{equation}~\label{eq:haldanemft}
    H_{V_2}^{\rm Fock} = V_2 \sum\limits_{\langle\langle i,j \rangle\rangle} \eta_{i, j} c_{i}^{\dagger}c_{j}^{} - \mu N,
\end{equation}
where $\eta_{i,j} = \langle c_j^{\dagger}c_i \rangle$ and we assume $\eta_{i,j}$ to be purely imaginary to mimic the HO. To maintain half-filling, here we set $\mu=0$. The HO is then quantified by the average magnitude of all $\eta_{i,j}$, given by
\begin{equation}~\label{eq:haldaneOP}
    \delta_{\rm HO}^{} = \frac{1}{N_{\langle\langle ... \rangle\rangle}}\sum\limits_{\langle\langle i,j \rangle\rangle} |\eta_{i,j}^{}|,
\end{equation}
where $N_{\langle\langle \dots \rangle\rangle}$ is the number of NNN hoppings contained in the system. In addition to calculating this order parameter self-consistently, we also monitor the total magnetic flux through the entire system to ensure that it remains negligible (typically $\lesssim 10^{-3}$).

Upon constructing the total effective single-particle Hamiltonian $\hat{h}_{\rm strain} + \hat{h}_{V_2}^{\rm Fock}$, where $\hat{h}_{\rm V_2}^{\rm Fock}$ is the matrix operator associated with $H_{V_2}^{\rm Fock}$ from Eq.~\eqref{eq:haldanemft}, and self-consistently solving for $\eta_{i,j}$, we obtain $\delta_{\rm HO}$ as a function of the NNN coupling strength $V_2$ for the $\{6,3\}$, $\{10,3\}$, and $\{14,3\}$ Dirac systems, shown in Figs.~\ref{fig:Haldane}(a), ~\ref{fig:Haldane}(e), and~\ref{fig:Haldane}(i), respectively. Next we discuss the results that are, however, qualitatively similar to the CDW order.

In all three systems, with a sufficiently strong axial magnetic field producing flat bands near zero energy, a non-trivial HO emerges at subcritical ($V_2 < V_{2,{\rm c}}$) interaction strengths. In the region of $V_2 \ll V_{2,{\rm c}}$, the magnitude of the HO is increased with a stronger axial magnetic field. However, when $V_2 \gg V_{2, {\rm c}}$, we observe that a stronger axial magnetic field reduces the magnitude of the HO. The origin of such contrasting scaling behavior of the HO with increasing axial magnetic field strength in these two regions is the same as the one discussed previously for the CDW order. Namely, in the weak coupling regime the near zero energy flat band is responsible for the nucleation of the HO, whose DOS increases with increasing $q$ yielding an amplified HO with increasing axial field. On the other hand, in the strong coupling regime the larger bandwidth with increasing axial field strength reduces the propensity toward the HO. Nonetheless, taken together, these observations demonstrate the axial magnetic catalysis of the HO at weak (subcritical) coupling, which was originally promoted from analytical arguments in Sec.~\ref{sec:axialmagcat}. We also note that due to the spatial separation of the near zero energy modes living on the $A$ and $B$ sublattices, otherwise living near the bulk and boundary of the system, respectively, the HO predominantly develops on the $A$ and $B$ sublattices in these two regions of the system. However, to ensure that the net magnetic flux through the entire system due to the HO is still zero (specifically $\lesssim 10^{-3}$) despite significant finite size effects, we repeat the whole computation by taking $q \to -q$ that reverses the direction of axial fields and the nature of the localization of near zero modes on two sublattices, and subsequently average the HO from these two configurations. In addition, we also make sure that in each iteration during the self-consistent calculations the real component of all the $\eta_{i,j}$ are sufficiently small ($\ll 10^{-4}$) so that they can be neglected totally.

\section{Hubbard repulsion and global antiferromagnet}~\label{sec:hubbard}

The two prior sections, discussing finite range Coulomb interactions, excluded the spin degrees of freedom. In this section, we take into account the on-site Hubbard repulsions between fermions with opposite spin projections that are living on the same site of the lattice, thereby necessitating a collection of spinful fermions. Such an interaction gives rise to an AFM order in half-filled bipartite lattices, represented by a staggered pattern of electronic spin between two complementary sublattices of the system~\cite{assaaurbach}. In order to include this interaction into our model, we consider the following interacting Hamiltonian
\begin{equation}
    H_U = U \sum\limits_i \left ( n_{i, \uparrow} - \frac{1}{2} \right ) \left ( n_{i, \downarrow} - \frac{1}{2} \right ) - \mu N,
\end{equation}
where $U(>0)$ is a tunable parameter controlling the strength of the on-site repulsion and $n_{i,\uparrow}$ ($n_{i,\downarrow}$) is the fermionic density operator acting on site $i$ with spin projection $\uparrow$ ($\downarrow$). Decomposing $H_U$ in the Hartree channel gives the effective single-particle Hamiltonian
\begin{equation}~\label{eq:hubbmft}
\begin{split}
    H_{U}^{\rm Har} &= U \sum\limits_{x\in\{A,B\}}\sum\limits_i \; \bigg[ \left( \langle n_{x,i}^{\uparrow} \rangle - \frac{1}{2} \right) \left( n_{x,i}^{\downarrow} - \frac{1}{2} \right)\\
    &+ \left( \langle n_{x,i}^{\downarrow} \rangle - \frac{1}{2} \right) \left( n_{x,i}^{\uparrow} - \frac{1}{2} \right) \bigg] - \mu N,
\end{split}
\end{equation}
where $\langle n_{A,i}^{\uparrow} \rangle$ ($\langle n_{B,i}^{\downarrow} \rangle$) is the average fermionic density on the $i$th site of the $A$ ($B$) sublattice with spin up (down) and to maintain the half-filling, we set $\mu=0$. Here, analogous to our treatment of the CDW order, we define $\delta_{A/B, i}^{\uparrow/\downarrow}$, which quantifies the deviations of a spin-dependent average electronic density from half-filling on site $i$ of the $A/B$ sublattice with spin $\uparrow / \downarrow$. We compute these quantities self-consistently by filling all the negative energy states of the effective single-particle Hamiltonian $\sigma_0 \otimes \hat{h}_{\rm strain} + \hat{h}^{\rm Har}_U$, where $\hat{h}^{\rm Har}_U$ is the matrix operator associated with Eq.~\eqref{eq:hubbmft}. In terms of these local-site order parameters, we then express the average fermionic density as
\begin{equation}
\langle n_{A,i}^{\sigma} \rangle= \frac{1}{2} + \sigma \; \delta_{A,i}^{\sigma}
\:\:\: \text{and} \:\:\:
\langle n_{B,i}^{\sigma} \rangle= \frac{1}{2} - \sigma \; \delta_{B,i}^{\sigma}, 
\end{equation}
where $\sigma=+$ and $-$, respectively correspond to $\uparrow$ and $\downarrow$ spin projections, and $\delta_A^\sigma$ ($\delta_B^\sigma$) is the average quantity over all $\delta_{A,i}^{\sigma}$ ($\delta_{B,i}^{\sigma}$). Furthermore, at half-filling, $\delta_{A,i}^{\sigma}$ and $\delta_{B,i}^{\sigma}$ must also adhere to the condition
\begin{equation}
\sum_{i \in A} \sum_{\sigma=\pm} \; \sigma \; \delta_{A}^{\sigma} - 
\sum_{i \in B}\sum_{\sigma=\pm} \; \sigma \; \delta_{B}^{\sigma} =0
\end{equation}
to maintain the half-filling condition.

\begin{figure}[t!]
\includegraphics[width=1.00\linewidth]{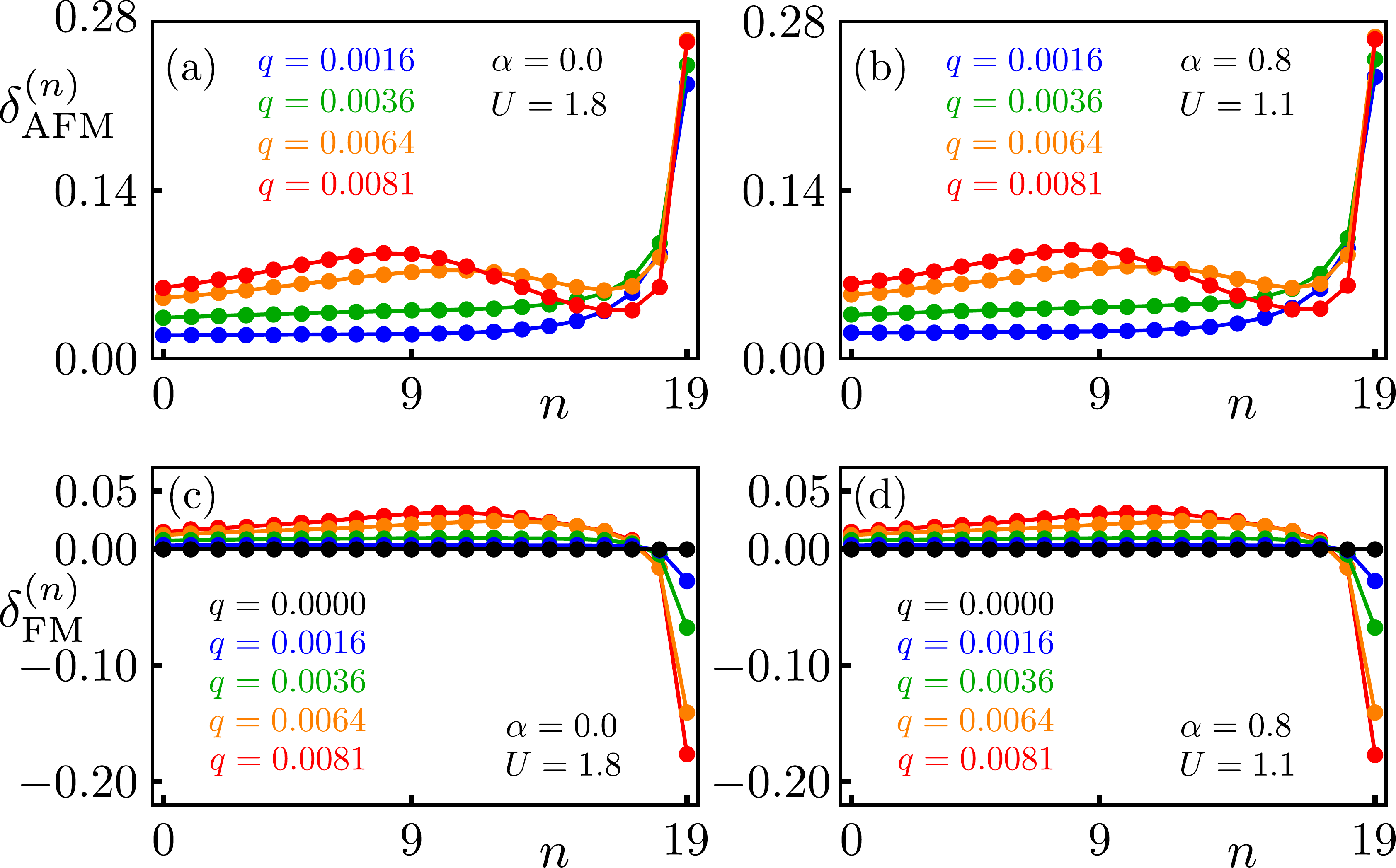}
\caption{Spatial variation of the self-consistent solutions for the local antiferromagnetic order ($\delta^{(n)}_{\rm AFM}$) for various strengths of the axial magnetic field (parameterized by $q$) with a fixed sub-critical strength of the on-site Hubbard repulsion $U$ (see Table~\ref{tab:criticalvalues}) on a (a) Hermitian ($\alpha=0$) and (b) non-Hermitian ($\alpha=0.8$) honeycomb lattice. The values of $q$ and $U$ are specified in the legends, and $n$ corresponds to the generation number of the honeycomb lattice (Sec.~\ref{sec:systemsize}). Notice that $\delta^{(n)}_{\rm AFM}$ is positive everywhere, yielding a net non-zero antiferromagnetic order ($\delta_{\rm AFM}$) when averaged over the entire system, see Fig.~\ref{fig:AFM}. Panels (c) and (d) are analogous to (a) and (b), respectively, for the same set of parameter values, but showing the spatial variation of the self-consistent solutions of the local ferromagnetic order ($\delta^{(n)}_{\rm FM}$). Notice that $\delta^{(n)}_{\rm FM}$, although finite everywhere, is of opposite signs in the bulk and near the boundary of the system, yielding a net zero magnetization in the whole system. We also show $\delta^{(n)}_{\rm FM}$ in the absence of any axial field ($q=0$) for which it is pinned at zero in the entire system. See Sec.~\ref{sec:hubbard} for a detailed discussion.   
}~\label{fig:GlobalAFM63}
\end{figure}

\begin{figure}[t!]
\includegraphics[width=1.00\linewidth]{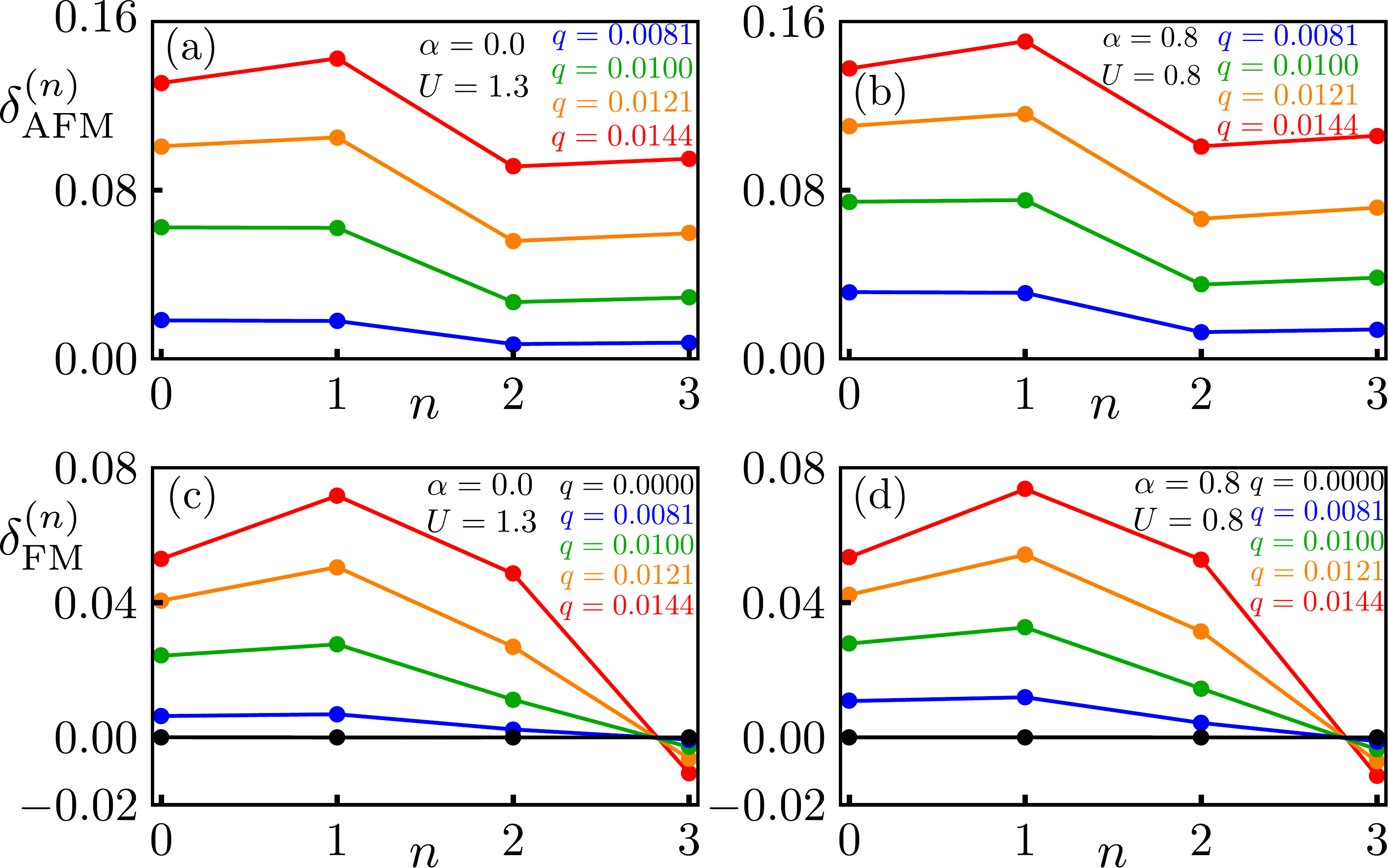}
\caption{Analogous to Fig.~\ref{fig:GlobalAFM63}, but for the $\{10,3\}$ hyperbolic Dirac system. Notice that the amplitude of $\delta^{(n)}_{\rm FM}$ near the boundary is much smaller in comparison to those in its own bulk and to those near the boundary of the honeycomb lattice. As hyperbolic lattices foster a large number of sites at the boundary, compared to that in its own bulk or to the number of sites on the edges of Euclidean lattices, the net magnetization in the whole system is still zero. All parameter values are mentioned in the legends. For details see Sec.~\ref{sec:hubbard}.   
}~\label{fig:GlobalAFM103}
\end{figure}

\begin{figure}[t!]
\includegraphics[width=1.00\linewidth]{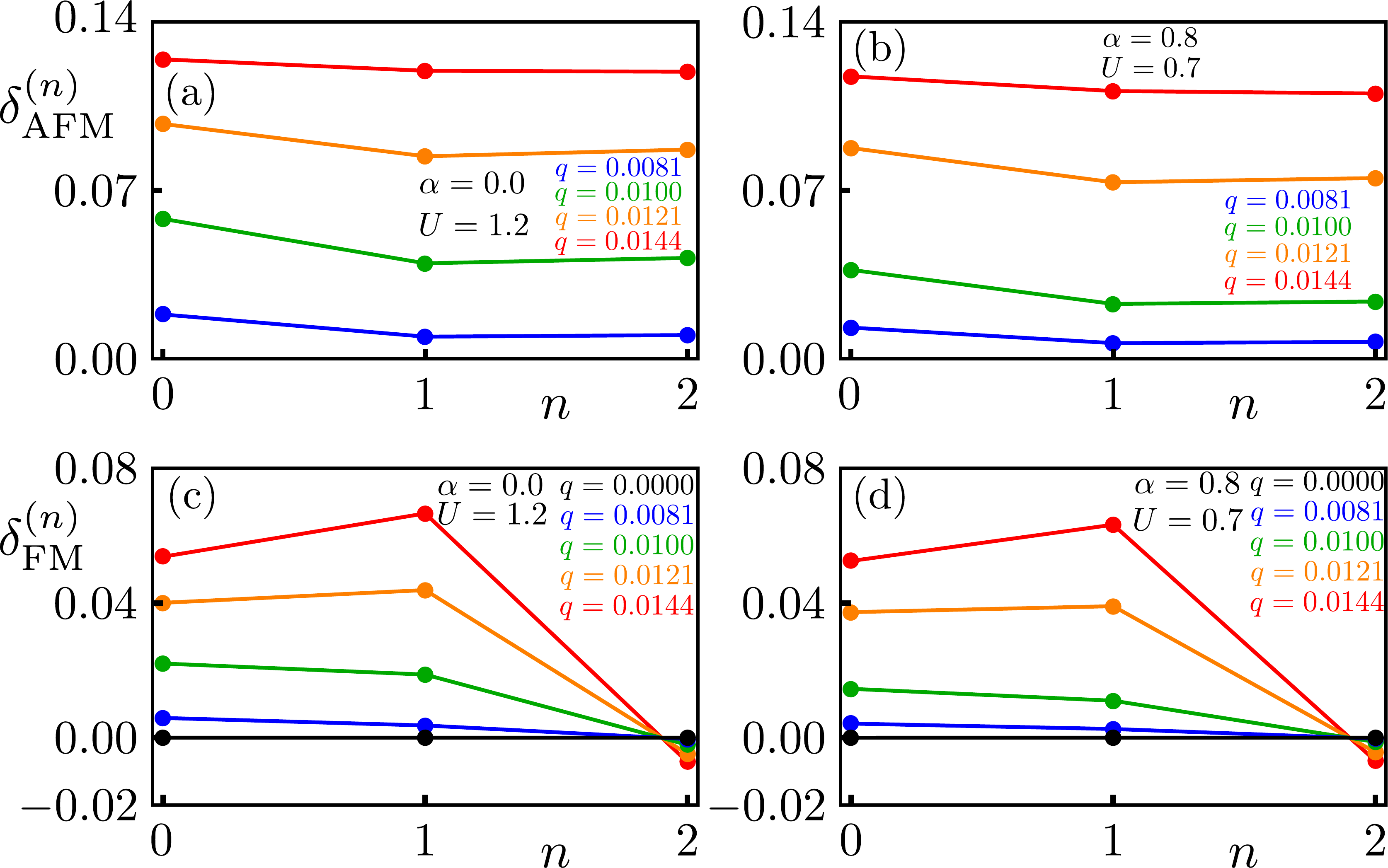}
\caption{Analogous to Figs.~\ref{fig:GlobalAFM63} and~\ref{fig:GlobalAFM103}, but for the $\{ 14, 3\}$ hyperbolic lattice.  All parameter values are mentioned in the legends. For details see Sec.~\ref{sec:hubbard}.   
}~\label{fig:GlobalAFM143}
\end{figure}

Equipped with $\delta_{A}^{\sigma}$ and $\delta_{B}^{\sigma}$, the order parameters for the AFM and FM orders immediately follow, and are respectively given by~\cite{AxialCataGra:5} 
\begin{eqnarray}
\delta_{\rm AFM} &=& (\delta_A^\uparrow - \delta_B^\uparrow - \delta_A^\downarrow + \delta_B^\downarrow)/2\\  
\text{and} \:\:\: \delta_{\rm FM} &=& (\delta_A^\uparrow + \delta_B^\uparrow - \delta_A^\downarrow - \delta_B^\downarrow)/2.
\end{eqnarray}
With the relevant order parameters defined, self-consistently calculate the AFM and FM order parameters. The results of this analysis are shown in Figs.~\ref{fig:AFM}(a),~\ref{fig:AFM}(e), and~\ref{fig:AFM}(i) for the strained $\{ 6,3 \}$, $\{ 10,3 \}$, and $\{ 14,3 \}$ lattices, respectively, but only for the AFM order, promoting its catalysis by axial magnetic fields for subcritical strengths of the on-site Hubbard repulsion. However, we note that in the presence of axial fields the Hubbard repulsion also allows for the FM order, which gives rise to phenomena that are absent in the former two cases with NN and NNN Coulomb repulsions. Such a competition between AFM and FM solely results from the spatial separation of the near zero energy modes localized on the $A$ and $B$ sublattices, residing in the bulk and near the boundary of the system, respectively. To scrutinize the competition between two such orders, we define the corresponding generation-localized order parameters $\delta_{\rm AFM}^{(n)}$ and $\delta_{\rm FM}^{(n)}$, which are calculated using the same methodology described above, while only accounting for sites on the $n$th generation (for definition of generation see Sec.~\ref{sec:systemsize}). In this way, we are able to track the local features of these two orders. The results are shown in Fig.~\ref{fig:GlobalAFM63} for the $\{6,3\}$ lattice, Fig.~\ref{fig:GlobalAFM103} for the $\{10,3\}$ lattice, and Fig.~\ref{fig:GlobalAFM143} for the $\{14,3\}$ lattice, all for a fixed subcritical coupling strength.

We also focus on the half-filled $\{ 6,3 \}$, $\{ 10,3 \}$, and $\{ 14,3 \}$ systems in the absence of any axial magnetic field, resulting in a zero FM order in each generation of the system. Conversely, in the presence of the field, the local FM order begins to take on a non-trivial texture, with positive FM order accumulating on the bulk generations. Despite this, the total FM order across the entire system remains pinned to zero (not shown explicitly) due to the development of a negative FM order near the edge. This cancellation of the FM orders by the edge contributions, along with the positive AFM order across all generations of the system, giving birth to a peculiar magnetic order which we name \emph{edge-compensated global AFM}. Such a outcome can be appreciated in the following way. The local FM order develops due to the existence of the near zero energy modes which live on complementary sublattices near the bulk and edge of the system (see Sec.~\ref{sec:zeromodes}), that simultaneously also yields a finite AFM order. However, the magnetization changes its sign where the near zero energy modes lose support in the bulk of the system, whereas the AFM order continue to survive with the same sign. The ground state in this way features a net zero magnetization, while fostering a finite AFM or Ne\'el order in the entire system.

All of these results hold true in the Euclidean $\{6,3\}$, and hyperbolic $\{10,3\}$ and $\{14,3\}$ lattice systems. However, it is worth noting that the edge-localized, negative FM orders in the hyperbolic systems are relatively much smaller than those appearing in the Euclidean honeycomb system. This difference is explained by the geometry of the hyperbolic lattices which, unlike their Euclidean counterparts, feature a majority of sites residing on the edge, even in the thermodynamic limit. As such, the small negative edge-localized FM order cancels with the one from the bulk and still yields a net zero FM order across the system, due to the large number of edge sites. Having established a unified picture of the axial magnetic catalysis in Euclidean and hyperbolic strained Dirac systems in the last three sections, next we proceed showcase its amplification in certain NH setups.

\section{NH Dirac fermions in axial magnetic fields}~\label{sec:NHconstruct}

Recently, it has been proposed that a certain class of coupling of a system with its environment (namely, a balanced gain and loss) can amplify the nucleation of spontaneous-symmetry breaking phases in certain channels, namely the commuting class masses which we define in a moment, by reducing the bandwidth of the free-fermion energy spectrum while maintaining its reality~\cite{biorthogonalQM:2}. The construction of the corresponding NH operator rests on the identification of an operator $\hat{h}_{\rm mass}$, that anticommutes with the original Hermitian model, which in this case is $\hat{h}_{\rm strain}$. Subsequently, an anti-Hermitian term $\hat{h}_{\rm mass}\hat{h}_{\rm strain}$ is recognized, giving way to the following desired NH operator~\cite{NHDirac:1, NHDirac:2, NHDirac:3, NHDirac:4, NHDirac:5}
\begin{equation}
    \hat{h}_{\rm NH} = \hat{h}_{\rm strain} + \alpha \hat{h}_{\rm mass}\hat{h}_{\rm strain},
    \label{eq:hNH}
\end{equation}
where $\alpha$ is a tunable real parameter controlling the non-Hermiticity, with $\alpha=0$ corresponding to the Hermitian limit. In this work, we choose
\begin{equation}~\label{eq:hmass}
    \hat{h}_{\rm mass} = \text{diag.}(\mathbf{I}, \mathbf{-I}),
\end{equation}
where $\mathbf{I}$ is an $N/2$-dimensional identity operator and the requisite condition $\{\hat{h}_{\rm strain}, \hat{h}_{\rm mass}\}=0$ can be explicitly verified by comparing with Eq.~\eqref{eq:hstrain}. With this choice of $\hat{h}_{\rm mass}$, the NH operator $\hat{h}_{\rm NH}$ takes the form
\begin{equation}
    \hat{h}_{\rm NH} = \begin{pmatrix}
        \mathbf{0} & (1+\alpha)\mathbf{t_{\rm strain}}\\
        (1-\alpha)\mathbf{t_{\rm strain}^{\top}} & \mathbf{0}
    \end{pmatrix},
    \label{eq:hNHspecific}
\end{equation}
which describes an imbalance in the inter-sublattice hopping amplitudes in opposite directions, shown in Fig.~\ref{fig:Geometry}(a)-(c). The spectrum of $\hat{h}_{\rm NH}$ is given by the set $\{E_{\rm NH}\} = \{E_{\rm strain}\}\sqrt{1-\alpha^2}$, where $\{E_{\rm strain}\}$ is the energy spectrum of the Hermitian operator $\hat{h}_{\rm strain}$. Note that this expression assumes spinless fermions, which can be straightforwardly extended for spinful fermions, yielding the corresponding NH operator $\sigma_0 \otimes \hat{h}_{\rm NH}$. All the eigenvalues are then two-fold Kramer's degenerate.

At this point, a few remarks on such a NH model are needed. First and foremost, notice that as long as $|\alpha|\leq1$, $\{E_{\rm NH}\}$ is all-real, thereby guaranteeing an infinite lifetime for all emergent quasiparticles and a well-defined filling factor in the open system. In this work we do not consider the parameter regime $|\alpha|>1$, yielding all imaginary eigenvalues in free-fermion systems, nor do we consider the singular cases $\alpha= \pm 1$, which represent the exceptional points where all eigenvalues coalesce to zero. Most importantly, note that the spectrum of the NH operator is simply a scaled version of the original Hermitian spectrum, with all eigenvalues pushed toward zero energy by a factor of $\sqrt{1-\alpha^2}$, thereby representing a compression of the bandwidth with increasing $\alpha$ as long as $|\alpha|<1$. This scaling behavior does not alter the characteristic linearly vanishing DOS near zero energy in the absence of axial fields, thereby leaving the Dirac classification invariant. The zero energy modes of the spectrum are left unaffected by the non-Hermiticity in the system, as well. These effects on the DOS, in systems with a strain-induced axial magnetic field, are shown in Fig.~\ref{fig:DOS} (bottom row). While the zeroth Landau level remains invariant under a change of the NH parameter $\alpha$, hyperbolic lattices show an increased peak height in the DOS at zero energy with a greater non-Hermiticity. Such a feature results from the lack of any Landau gap in hyperbolic systems (Sec.~\ref{sec:zeromodes}), resulting in the presence of near zero energy modes which are pushed into the energy bin centered at zero energy, used to compute the DOS around it.

\section{NH Amplification}~\label{sec:NHamplify}

In Secs.~\ref{sec:coulomb}-\ref{sec:hubbard}, we studied orders induced by the NN and NNN Coulomb interactions, as well as those arising from on-site Hubbard interactions, see Eqs.~\eqref{eq:massoperators} and~\eqref{eq:haldaneorder}. Notably, the resulting CDW, HO, and AFM masses commute with $\sigma_0 \otimes \hat{h}_{\rm mass}$ [Eq.~\eqref{eq:hmass}], meaning that they still serve as Dirac masses for the NH model as all of them fully anticommute with $\sigma_0 \otimes \hat{h}_{\rm NH}$. We call members of such a family of orders the \emph{commuting class masses} (due to their commutation relation with $\hat{h}_{\rm mass}$).

To study such mass orders in our NH model, we decompose the corresponding four-fermion interactions in the Hartree or Fock channels to obtain effective single-particle descriptions $\hat{h}^{\rm MF}_{\rm int}$, where $\hat{h}^{\rm MF}_{\rm int}=\hat{h}^{\rm Har}_{V_1}$ or $\hat{h}^{\rm Fock}_{V_2}$ or $\hat{h}^{\rm Har}_{U}$ (Secs.~\ref{sec:coulomb}-\ref{sec:hubbard}). Subsequently, we form the mean field Hamiltonian $\hat{h}_{\rm NH} + \hat{h}^{\rm Har/Fock}_{\rm int}$, whose energy spectrum is given by $E_{{\rm NH},i}^{\Delta} = \sqrt{E_{i}^2(1-\alpha^2) + \Delta^2}$ for all $i$. Defining the free energy analogous to Eq.~\eqref{eq:freeenergy} and minimizing with respect to $\Delta$, we arrive at the gap equation in NH systems
\begin{equation}
    \frac{1}{g} = \frac{D_0}{|\Delta|} + {\sum\limits_{i=1,2,\cdots}}^\prime \frac{D_i}{\sqrt{E_i^2(1-\alpha^2) + \Delta^2}}.
    \label{eq:freeenergyminNH}
\end{equation}
The first term results from the split zero-energy states, playing no role in NH amplification, which instead solely results from the second term. To see the amplification of the axial magnetic catalysis by the non-Hermiticity, consider the case where $\Delta$ is fixed. In this case, increasing the value of $\alpha$ results in a smaller $g$, thereby demonstrating the ability for a smaller coupling strength to open a gap with equal magnitude in the system due to the introduction of non-Hermiticity. Alternatively, one can also fix $g$, and vary $\alpha$ and $\Delta$, the result of which is a larger gap at a given coupling strength when non-Hermiticity is increased. Together, these arguments demonstrate the NH amplification of the axial magnetic catalysis that is operative for CDW order, HO, and AFM order.

Note that, the NNN interactions, that give rise to a HO, does not perfectly anticommute with the free-fermion model due to the finiteness of the systems, as stated in Sec.~\ref{sec:axialmagcat}. However, like in the case of the strained Hermitian system, this does not preclude the above free energy argument from justifying the NH amplification of the axial magnetic catalysis of the HO, with one important note. Since a perfect anticommutation relation is not strictly enforced, there is no analytical argument that guarantees an entirely real energy spectrum of the corresponding effective single-particle Hamiltonian obtained after decomposing the NNN Coulomb repulsion in the Fock channel. Therefore, as we perform this computation, we must monitor the eigenvalues to ensure that the imaginary components remain negligible relative to their real component counterparts (by a factor of $\lesssim 10^{-3}$).

\section{Biorthogonal quantum mechanics}~\label{sec:biorthogonal}

Since we embark to study non-Hermitian models of strained Dirac systems, a brief review of the biorthogonal quantum mechanics is warranted here. To begin, define a NH operator $\mathcal{H}$, meaning that $\mathcal{H}\neq\mathcal{H}^\dagger$. Then, the left and right eigenvector solutions of this operator, for a given eigenvalue $E$, are given by
\begin{equation}
    \mathcal{H}\ket{E^{\rm R}} = E\ket{E^{\rm R}}\;\; \text{ and }\;\; \bra{E^{\rm L}}\mathcal{H} = \bra{E^{\rm L}}E,
    \label{eq:eigenprob}
\end{equation}
respectively, where the left (right) eigenvector is labeled by the superscript $\rm L$ ($\rm R$), which is necessitated by the fact they are no longer related by a Hermitian adjoint operation. In our models, the effective single-particle Hamiltonian plays the role of $\mathcal{H}$ and $E$ is real. The left and right eigenvectors of $\mathcal{H}$ now satisfy the bi-orthonormality condition
\begin{equation}
    \bra{E^{\rm L}_m}\ket{E^{\rm R}_n} = \delta_{m,n},
    \label{eq:biorthonorm}
\end{equation}
where $\bra{E^{\rm L}_m}$ ($\ket{E^{\rm R}_n}$) is the left (right) eigenvector associated with eigenvalue $E_m$ ($E_n$) and $\delta_{m, n}$ is the Kronecker delta symbol~\cite{biorthogonalQM:1}.

The bi-orthonormality condition can now be extended to define the probability of observing a state with a given eigenvalue $E$ on the $i$th site of the lattice, giving~\cite{biorthogonalQM:2}
\begin{equation}
    P_i = \bra{E^{\rm L}}\ket{i} \bra{i}\ket{E^{\rm R}}
    \label{eq:biorthoprob}
\end{equation}
where $\ket{i}$ is the site-localized Wannier function centered on site $i$. This definition is utilized in the self-consistent calculation of the CDW ($\delta_{\rm CDW}$), and AFM ($\delta_{\rm AFM}$) and FM ($\delta_{\rm FM}$) order parameters with NN Coulomb and on-site Hubbard repulsions, respectively. Namely, after diagonalizing the mean-field Hamiltonian, we take all the eigenvectors associated with the filled Dirac sea ($E<0$) and use them to compute the average fermionic density (spin independent or dependent) on each site of the lattice. These densities enter the effective single-particle Hamiltonian of the parent interacting models [see Eqs.~\eqref{eq:coulombmft} and~\eqref{eq:hubbmft}], which in turn defines the corresponding order parameters (see Secs.~\ref{sec:coulomb} and~\ref{sec:hubbard}). For the Haldane currents, we compute
\begin{equation}
    \eta_{i,j} = \langle c_j^{\dagger}c_i^{} \rangle = {\sum\limits_E}^\prime \bra{E^{\rm L}}\ket{j}\bra{i}\ket{E^{\rm R}},
\end{equation}
where the sum is over all energies $E$ in the filled Dirac sea with $E<0$, which enters the effective single-particle Hamiltonian from Eq.~\eqref{eq:haldanemft}, obtained after decomposing NNN Coulomb repulsion in the Fock channel and defines the corresponding order parameter $\delta_{\rm HO}$ [see Eq.~\eqref{eq:haldaneOP}].

\begin{figure}[t!]
\includegraphics[width=1.00\linewidth]{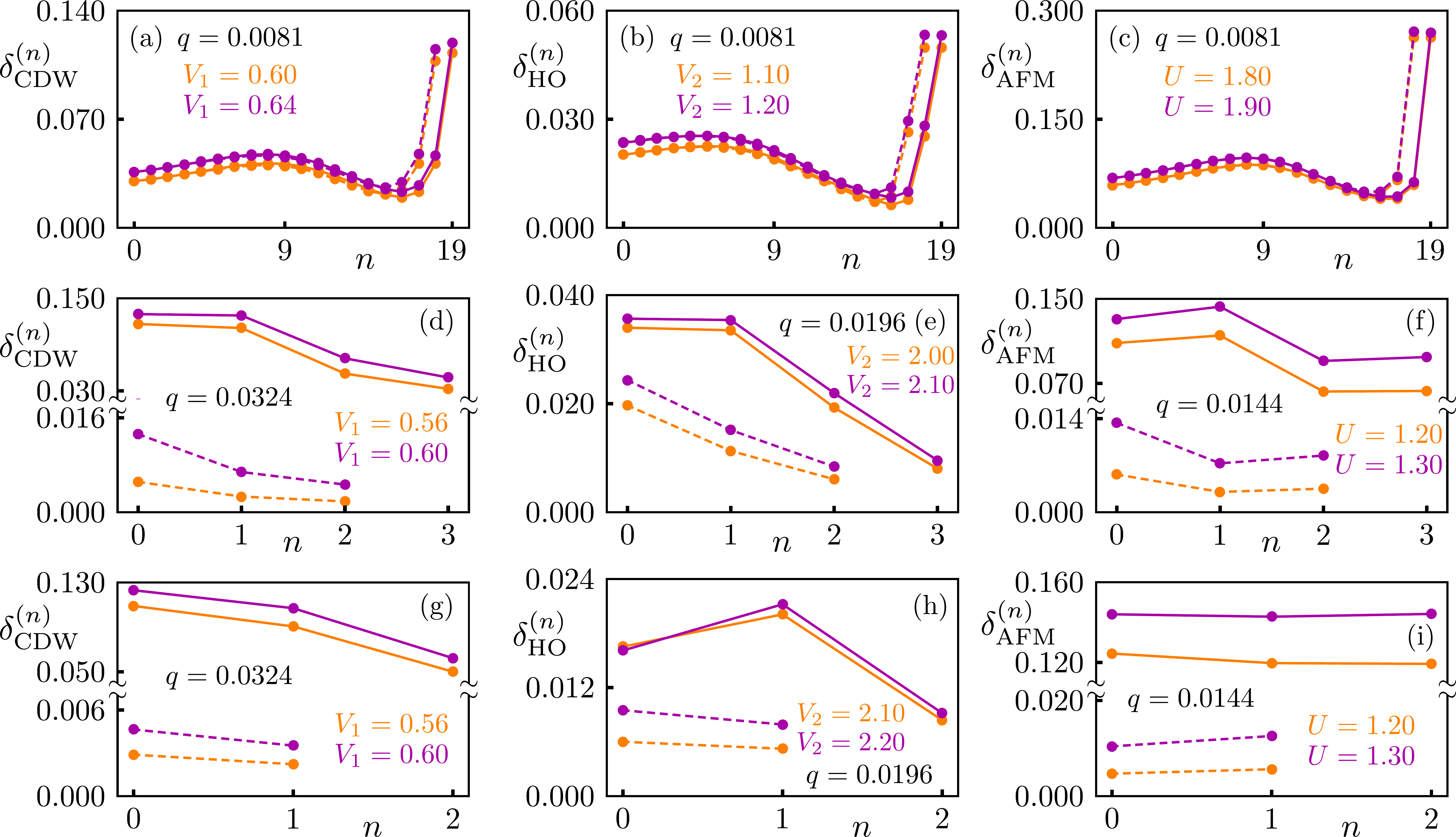}
\caption{(a) Dependence of the local charge density wave order parameter $\delta^{(n)}_{\rm CDW}$ on the generation number $n$ of the honeycomb lattice with two different system sizes for a fixed strength of the axial magnetic field (set by $q$) and two different subcritical values of the nearest-neighbor Coulomb repulsion ($V_1$) among spinless fermions. Throughout, the results for two different interaction strengths (system sizes) are shown in different colors (line styles). (b) is the same as (a), but showing the generation ($n$) dependence of the local Haldane order $\delta^{(n)}_{\rm HO}$ for two subcritical values of the next-nearest-neighbor Coulomb repulsion ($V_2$) among spinless fermions for a fixed $q$ (axial magnetic field) on honeycomb lattices of two different sizes. (c) is the same as (a) and (b), but displaying the generation ($n$) dependence of the local antiferromagnetic order $\delta^{(n)}_{\rm AFM}$ for two subcritical values of the on-site Hubbard repulsion ($U$) among spinful fermions for a fixed $q$ (axial field) on honeycomb lattices of two different sizes. Panels (d)-(f) [(g)-(i)] are the same as subfigures (a)-(c), respectively, but on $\{ 10, 3\}$ [$\{ 14, 3\}$] hyperbolic lattices of two different sizes. All the results are shown for Hermitian systems with $\alpha=0$. All the parameter values are mentioned in the legends. See Sec.~\ref{sec:finitesize} for details and Table~\ref{tab:criticalvalues} for critical values of the various interactions for the corresponding orders.   
}~\label{fig:finsizeHer}
\end{figure}

A brief note on numerical eigensolvers is due at this stage. While eigensolvers designed for Hermitian operators are highly stable and robust, owing to simplifying assumptions guaranteed by the Hermiticity, thereby ensuring the orthonormality condition, general eigensolvers (employed here to diagonalize a NH operator) do not enjoy such advantages, often leading to the violation of the biorthonormality condition quoted in Eq.~\eqref{eq:biorthonorm}. Namely, when using these general solvers to obtain eigenvectors of a NH operator, the biorthonormality condition, may be violated even when the operator hosts a non-degenerate or a nearly degenerate spectrum, sourced by the eigenvalues that are distinct but lie close to one another. Thus, in order to ensure that the eigensolver can distinguish between these nearly degenerate eigenvalues, we add a small amount of on-site disorder to the system. The disorder is sampled from a uniform box distribution $[-W/2,W/2]$ with $W \sim 10^{-3}$ typically. To minimize the shift in the Fermi energy due to even such small disorder, we subtract a uniform background average of all sampled disorder from each lattice site. We also ensured that the self-consistent solutions for various orders is insensitive to the disorder configuration for a fixed $W$.

\section{NH interacting systems}~\label{sec:NHinteract}

Having reviewed the requisite tools of the biorthogonal quantum mechanics, we now numerically demonstrate the NH amplification of axial magnetic catalysis, which we have already promoted using analytical arguments in Sec.~\ref{sec:NHamplify}. To this end, we consider the CDW order, HO, and AFM and FM orders, arising from the NN Coulomb, NNN Coulomb, and on-site Hubbard interactions, respectively. These results are obtained using analogous methods to those employed in Secs.~\ref{sec:coulomb}-\ref{sec:hubbard}, after adapting the proper biorthogonal definitions of the corresponding order parameters. The numerical results are shown in Fig.~\ref{fig:CDW} for the CDW order, Fig.~\ref{fig:Haldane} for the HO, and Fig.~\ref{fig:AFM} for the AFM order. Additionally, we study the generation-localized order parameters for the competing AFM and FM orders in Figs.~\ref{fig:GlobalAFM63},~\ref{fig:GlobalAFM103}, and~\ref{fig:GlobalAFM143} for the $\{6,3\}$, $\{10,3\}$, and $\{14,3\}$ lattices, respectively, suggesting that edge-compensated global antiferromagnet continues to be the ground state of the Hubbard model for strained NH Euclidean and hyperbolic Dirac fermions.

For all the orderings belonging to the commuting class mass family, we observe the emergence of the respective order at subcritical interaction strengths (see Table~\ref{tab:criticalvalues}) in systems with a sufficiently strong axial magnetic field, with a stronger field strength corresponding with an increasing magnitude of the order. Furthermore, we also note that the reversal of this trend at stronger interaction strengths also shows up in the NH systems, as previously noted for the Hermitian case. Nonetheless, these observations verify that the axial magnetic catalysis mechanism remains operative in NH strained Dirac systems.

We now seek to elucidate the contribution of the NH degree of freedom by considering the $\{6,3\}$, $\{10,3\}$, and $\{14,3\}$ systems, all with a fixed strength of the axial magnetic field and over a range of subcritical and above-critical coupling strengths (see Table~\ref{tab:criticalvalues}) of each of the three local interactions, considered in this work. This analysis is shown in the rightmost columns of Fig.~\ref{fig:CDW} for the CDW order, Fig.~\ref{fig:Haldane} for the HO, and Fig.~\ref{fig:AFM} for the AFM order. In all cases, we observe an increase in the ordering amplitudes with increasing non-Hermiticity as long as $|\alpha|<1$, thereby demonstrating the NH amplification of the order, catalyzed by the axial magnetic field.

\begin{figure}[t!]
\includegraphics[width=1.00\linewidth]{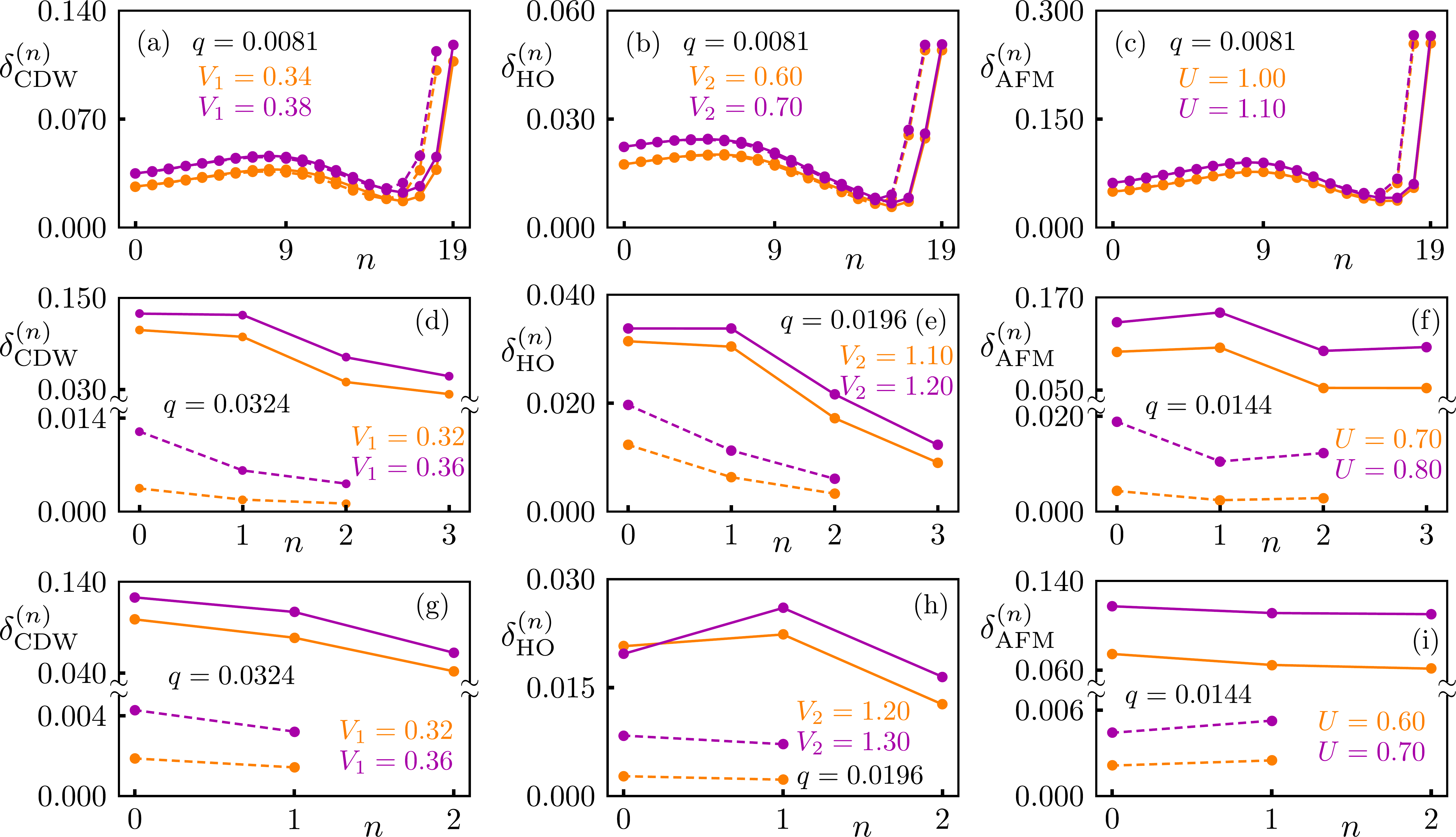}
\caption{Same as Fig.~\ref{fig:finsizeHer} but in non-Hermitian systems with $\alpha=0.8$. See Sec.~\ref{sec:finitesize} for details and Table~\ref{tab:criticalvalues} for the critical interaction strengths for the corresponding orders.
}~\label{fig:finsizeNH}
\end{figure}

\section{Finite size analysis}~\label{sec:finitesize}

The preceding sections were dedicated to discussing analytical and numerical evidence of the axial magnetic catalysis. While the former approach is devoid of any finite-size effects, the latter one is obtained from self-consistent solutions of various orders on lattices with open boundary conditions. As a result, it is natural to ask whether our numerical results remain operative in the thermodynamic limit, as the total number of generations $N \rightarrow \infty$. In order to address this question, here we study the the generation-dependent CDW ($\delta_{\rm CDW}^{(n)}$), Haldane ($\delta_{\rm HO}^{(n)}$), and AFM ($\delta_{\rm AFM}^{(n)}$) order parameters on strained Dirac systems with $N$ and $N-1$ total number of generations. Specifically, we consider honeycomb or $\{6,3\}$ lattices with $18$ and $19$ generations, respectively hosting $2166$ and $2400$ sites, as well as hyperbolic $\{10,3\}$ ($\{14,3\}$) lattices with $2$ ($1$) and $3$ ($2$) number of generations, hosting $490$ ($168$) and $2880$ ($1694$) sites, respectively. The results of this analysis are shown in Fig.~\ref{fig:finsizeHer} for the Hermitian systems with a finite fixed uniform axial magnetic field and two subcritical strengths of the relevant interaction.

Notice that in all cases, the larger system displays a greater or approximately equal order on each generation. This is a consequence of adding an extra generation to the smaller system, which introduces additional axial magnetic flux through the system and, following from index theorems, an increased number of zero energy modes. Furthermore, the hyperbolic systems, when compared to their Euclidean counterpart, noticeably show a greater jump in the magnitude of the order parameter with the inclusion of an extra generation. This is the consequence of the geometry of the hyperbolic lattice, where each added generation results in an abrupt increase in the number of plaquettes in the system, enclosing the axial fields. These findings, extended to the thermodynamic limit, suggest that an axial magnetic field does bring about a finite ordering in the entire system, despite being at a subcritical interaction strength. The conclusions drawn from extrapolating to the thermodynamic limit remain the same (qualitatively) in NH systems, as displayed in Fig.~\ref{fig:finsizeNH}. In conclusion, these results provide confidence that our numerical results indeed reflect an (NH amplified) axial magnetic catalysis mechanism that is operative on strained Dirac systems in both flat Euclidean and negatively curved hyperbolic spaces, in the thermodynamic limit.

\section{Summary and discussion}~\label{sec:summarydiscussion}

To summarize, here we show that $5$-fold and $7$-fold rotational symmetry preserving hopping modulations can mimic the effect of external strain on $\{10,3 \}$ and $\{14,3\}$ hyperbolic Dirac lattices, respectively, producing ${\mathcal T}$-preserving axial magnetic fields. Coupling between massless Dirac fermions, living on a negatively curved hyperbolic space, and the axial magnetic field produces a flat band near the zero energy. Formation of such an axial flat band can be conducive to the nucleation of various ordered phases even for sufficiently weak (subcritical) local four-fermion interactions, thereby presenting a realistic and experimentally viable route to trigger spontaneous symmetry breaking even in weakly correlated hyperbolic Dirac fluids. In particular, weak NN (NNN) Coulomb repulsion favors the CDW (HO), while a  weak on-site Hubbard repulsion supports the edge-compensated global AFM order. Throughout, we draw parallels between these findings and the similar ones, previously reported in the literature for tri-axially strained graphene's honeycomb lattice (symmetric under $3$-fold rotations)~\cite{index:3, AxialCataGra:1, AxialCataGra:2, AxialCataGra:3, AxialCataGra:4, AxialCataGra:5, AxialCataGra:6}, thereby unifying the landscape of emergent phenomena, setting in via the axial magnetic catalysis in strained Euclidean and hyperbolic Dirac liquids. Finally, we show that axial magnetic catalysis can be amplified with certain types of non-Hermiticity in the system that, for a specific exemplary realization, yields an imbalance in the hopping amplitudes between complementary sublattices in the opposite directions, particularly when all the eigenvalues in the absence of local four-fermion interactions are guaranteed to be real.

A tempting future extension of the current pursuit would be to investigate the possible superconducting instabilities of axial flat bands when electronic interaction acquires a net small attractive component or when an attractive interaction is sourced by electron-phonon interactions~\cite{axialSC:1, axialSC:2, axialSC:3}. On graphene's honeycomb lattice, the strain-induced axial flat band can support pure $s$-wave and $f$-wave pairings, as well as a ${\mathcal T}$-breaking $f+is$ paired state~\cite{axialSC:2}. On the other hand, the pairing symmetry of the superconducting ground state within the zero energy axial flat band in strained hyperbolic Dirac systems remains an open problem, at this stage.     
\begin{figure*}[t!]
\includegraphics[width=0.85\linewidth]{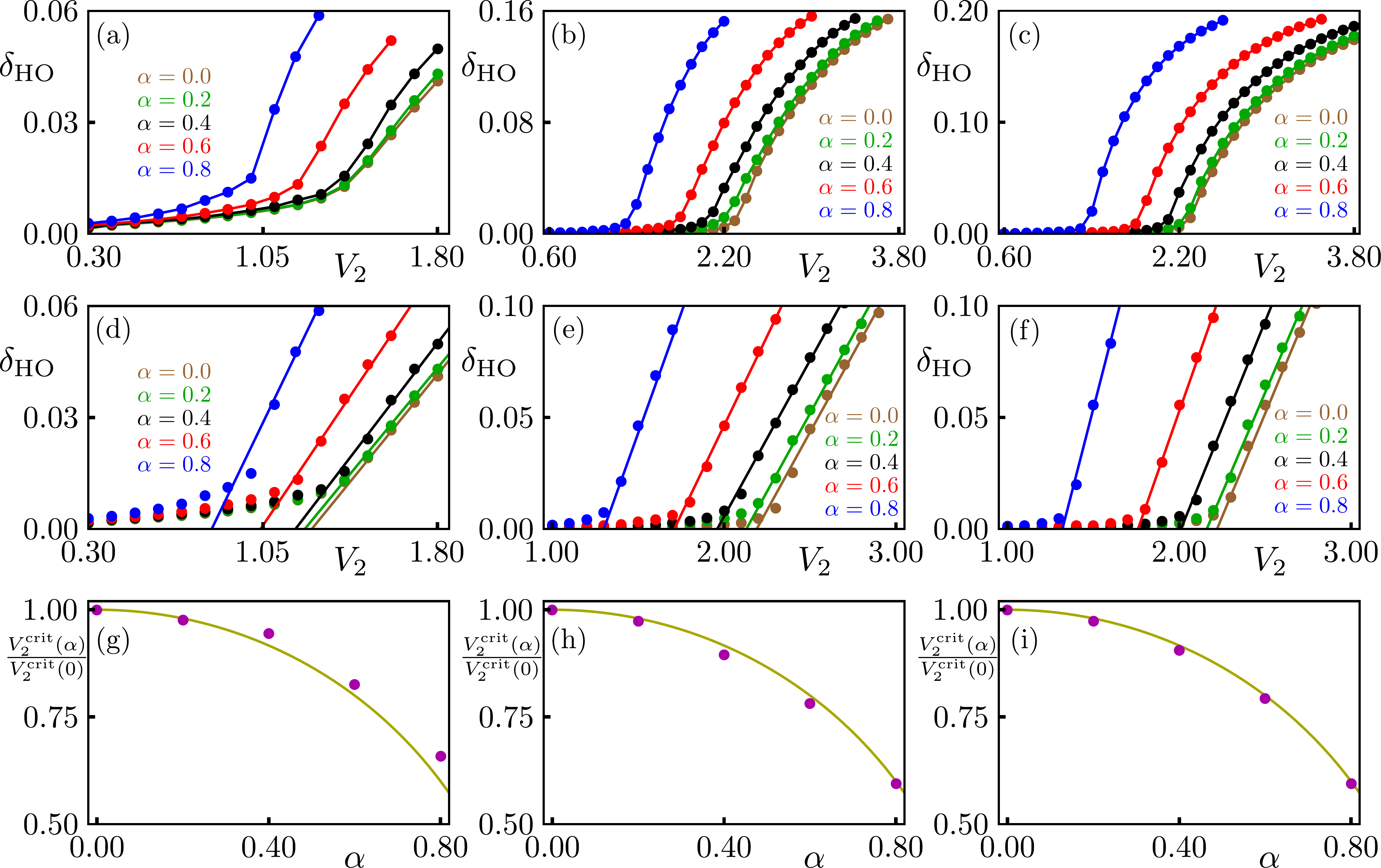}
\caption{Self-consistent solutions of the Haldane order ($\delta_{\rm HO}$), averaged over the entire system, as a function of the next-nearest-neighbor Coulomb repulsion ($V_2$) in the absence of any strain or axial magnetic field on (a) an Euclidean honeycomb or $\{ 6,3 \}$ lattice, and (b) $\{ 10, 3\}$ and (c) $\{ 14,3 \}$ hyperbolic lattices for various choices of the non-Hermitian parameter $\alpha$ that includes the Hermitian limit ($\alpha=0$). Computation of the critical strength of the next-nearest-neighbor Coulomb repulsion $V_{2, {\rm c}}^{}(\alpha)$ for the Haldane order, obtained from the linear fit of $\delta_{\rm HO}$ with $V_2$ around the critical one [see Eq.~\eqref{eq:linearfitcritical}], on (d) an Euclidean honeycomb or $\{ 6,3 \}$ lattice, and (e) $\{ 10, 3 \}$ and (f) $\{ 14, 3 \}$ hyperbolic lattices, which are also quoted in Table~\ref{tab:criticalvalues}. The ratio $V^{}_{2, {\rm c}}(\alpha)/V^{}_{2, {\rm c}}(0)$ as a function of $\alpha$ on (g) an Euclidean honeycomb or $\{ 6,3 \}$ lattice, and (h) $\{ 10,3 \}$ and (i) $\{ 14,3 \}$ hyperbolic lattices, where isolated dots correspond to the values obtained from numerical analyses and solid lines represent the theoretical predictions of this ratio being equal to $\sqrt{1-\alpha^2}$, see Eq.~\eqref{eq:scalingNHcritical}.   
For details see Appendix.~\ref{app:nostrainhaldane}.   
}~\label{fig:HaldaneNoAxial}
\end{figure*}

The central challenge for any experimental verification of our concrete theoretical predictions stems from the difficulty in creating quantum hyperbolic lattices and subsequently emulating the strain-induced modulated hopping pattern therein to harness axial fields therein. Still, designer quantum materials and optical lattices constitute two promising platforms for a laboratory-based demonstration of our results. Encouragingly, a tri-axial strain-induced axial magnetic field has already been realized on designer graphene~\cite{StrainExpGra:3}. On such a platform, $\{ 10,3\}$ and $\{ 14, 3\}$ hyperbolic tessellations can, for example, be created by growing its substrate on a suitable material with a different thermal expansion coefficient at high temperature. The desired negatively-curved substrate can be realized on such a heterostructure under cooling, which then can be decorated by the sites of hyperbolic lattices. Subsequently, the requisite $5$-fold and $7$-fold rotational symmetric spatially modulated hopping amplitudes on these two designer hyperbolic lattices, respectively, can be engineered by adequately tuning the distance between atoms on various sites, as previously employed to create tri-axially strained designer graphene~\cite{StrainExpGra:3}. In cold atomic setups, on which a honeycomb lattice has already been emulated~\cite{optical:1}, hyperbolic tessellations can possibly be achieved by suitable arrangements of the laser traps or optical tweezers. On such a highly tunable platform, axial magnetic fields can be tailored through spatial modulations of laser traps, for example, yielding the desired spatially modulated hopping patterns, shown in Fig.~\ref{fig:Geometry}. While on optical lattices the strength of the Hubbard repulsion can be tuned to a large degree~\cite{optical:1, optical:2, optical:3} to observe some of the correlated phases we discussed in this work, designer materials have also recently become quite promising in this direction~\cite{DesignerDiracCrit}.

Some classical metamaterials are also promising platforms for emulating the operator associated with single-particle Hamiltonian, capturing the effect of strain or axial fields, namely $\hat{h}_{\rm strain}$ therein. For example, hyperbolic lattices have been realized on topolectric circuits~\cite{metaHL:1} and photonic lattices~\cite{metaHL:2} recently, where the desired hopping modulations shown in Fig.~\ref{fig:Geometry} can be introduced to showcase the emergence of an axial field-induced flat band near zero energy. Fascinatingly, arrays of coupled microwave resonators serves as yet another prominent metamaterial platform where hyperbolic tessellations and their desired deformation, mimicking the axial field, can be realized to observe topological flat bands near zero energy, which has recently been demonstrated for tri-axially strained meta-graphene~\cite{metaHL:3}.

\acknowledgments 

This work was supported by the NSF CAREER Grant No.\ DMR-2238679 of B.R. Portions of this research were conducted on Lehigh University’s Research Computing infrastructure that is partially supported by the NSF Award No.\ 2019035.

\appendix

\section{Haldane order in the absence of strain}~\label{app:nostrainhaldane}

In a previous work~\cite{biorthogonalQM:2}, we determined the critical strengths of the NN Coulomb and on-site Hubbard repulsions, yielding CDW and AFM orders, respectively, in strain-free honeycomb, $\{10, 3\}$, and $\{14, 3\}$ lattice systems with different values of the non-Hermiticity ($\alpha$), which are quoted in Table~\ref{tab:criticalvalues}. In this appendix, we extend this analysis to determine the critical values of the NNN Coulomb repulsion ($V_{2,c}$), yielding the HO, which are self-consistently obtained using the methodology outlined in Secs.~\ref{sec:haldane} and \ref{sec:biorthogonal}. The results are shown in Fig.~\ref{fig:HaldaneNoAxial}(a),~\ref{fig:HaldaneNoAxial}(b), and~\ref{fig:HaldaneNoAxial}(c) for $\{6,3\}$, $\{10,3\}$, and $\{14,3\}$ lattices, respectively. Consistent with their Dirac classifications, we observe a significant non-trivial solution of HO only once a critical value of the NNN Coulomb interaction is surpassed. In order to identify this critical value, we rely on the mean field gap equation given by
\begin{equation}
    \frac{1}{V_2^{}} = \int\limits_0^{E_{\rm cut}(\alpha)} \frac{\rho(E, \alpha)}{\sqrt{E^2 + \Delta^2}}{\rm d}E(\alpha),
    \label{eq:appmftgap}
\end{equation}
where $E_{\rm cut}(\alpha) = E_{\rm cut}(0)\sqrt{1-\alpha^2}$ is the $\alpha$-dependent high-energy cutoff up to which the DOS is given by $\rho (E, \alpha) = |E|/v^2_{\rm F}(\alpha)$ with $v_{\rm F}(\alpha)$ as $\alpha$-dependent Fermi velocity, given by $v_{\rm F}(\alpha)=v_{\rm F}(0)\sqrt{1-\alpha^2}$, and $\Delta$ is the magnitude of the HO. Note that the above gap equation is the continuum analogue of Eq.~\eqref{eq:freeenergyminNH}, yielding
\begin{equation}
    \frac{1}{V_2^{}} = \frac{1}{v^2_{\rm F}(\alpha)} \: \left( \sqrt{E_{\rm cut}(\alpha)^2 + \Delta^2} - |\Delta| \right).
\end{equation}
The critical coupling constant is given by $V_{2, {\rm c}}= v^2_{\rm F}(\alpha)/|E_{\rm cut}(\alpha)|$ (obtained by setting $\Delta=0$ in the last equation), which also depends on the NH parameter $\alpha$, about which more in a moment. Substituting $V_{2, {\rm c}}$ in the last equation, while restricting ourselves to the regime where $\Delta/E_{\rm cut}(\alpha) \ll 1$, we obtain
\begin{equation}~\label{eq:linearfitcritical}
    |\Delta| \approx \frac{v^2_{\rm F}(\alpha)}{V_2 V_{2,c}} \: \left( V_2^{} - V_{2, {\rm c}} \right).
\end{equation}
We use such a linear growth of the HO once it starts to rise considerably from near the trivial values to determine the corresponding critical NNN repulsion, quoted in Table~\ref{tab:criticalvalues}. Corresponding analyses are shown in Fig.~\ref{fig:HaldaneNoAxial}(d),~\ref{fig:HaldaneNoAxial}(e), and~\ref{fig:HaldaneNoAxial}(f)  for $\{6,3\}$, $\{10,3\}$, and $\{14,3\}$ lattices, respectively. Our estimated value of $V_{2,c}$ on honeycomb lattice for $\alpha=0$ agrees reasonably with the previous estimation~\cite{NNNGra:MF}.

Finally, we note that due to the reduction of the bandwidth resulting from the non-Hermiticity in the system (parameterized by $\alpha$), the critical strength of the NNN repulsion in pristine Dirac systems (no axial fields) can be reduced, which follows the scaling law
\begin{equation}~\label{eq:scalingNHcritical}
    V_{2, {\rm c}}^{}(\alpha) = \frac{v^2_{\rm F}(\alpha)}{E_{\rm cut}(\alpha)}= \frac{v^2_{\rm F}(0)(1-\alpha^2)}{E_{\rm cut}(0) \sqrt{1-\alpha^2}}= V_{2,c}(0)\sqrt{1-\alpha^2}.
\end{equation}
This expected fit function for $V_{2, {\rm c}}(\alpha) / V_{2, {\rm c}}(0)$ is shown in Fig.~\ref{fig:HaldaneNoAxial} (g) for the $\{6,3\}$ lattice, Fig.~\ref{fig:HaldaneNoAxial} (h) for the $\{10,3\}$ lattice, and Fig.~\ref{fig:HaldaneNoAxial} (i) for the $\{14,3\}$ lattice.


\end{document}